\documentclass[aps,prl,reprint,superscriptaddress,longbibliography]{revtex4-2}
\usepackage[T1]{fontenc}
\usepackage{amsmath,amssymb,mathtools}
\usepackage{bm}
\usepackage{graphicx}
\usepackage{booktabs}
\usepackage{microtype}
\usepackage{xcolor}
\usepackage{xurl}
\usepackage{hyperref}
\definecolor{PRLPurple}{HTML}{2E3092}
\hypersetup{colorlinks=true,allcolors=PRLPurple}
\graphicspath{{./}}
\allowdisplaybreaks[0]
\newcommand{\TcL}{T_c^{\mathrm{1L}}}

\newcommand{\intk}{\int\!\frac{d^2k}{(2\pi)^2}}
\newcommand{\intq}{\int_{|\mathbf q|<\Lambda}\!\frac{d^2\mathbf q}{(2\pi)^2}}
\hypersetup{pdftitle={Quantum phase fluctuations quench chiral superconductivity in twisted overdoped cuprate bilayers},pdfauthor={Yin Shi, Mengxian Zhao, Fei Yang, Miao Liu, Sheng Meng}}
\input{arxiv-layout}
\begin{document}
\title{Quantum phase fluctuations quench chiral superconductivity in twisted overdoped cuprate bilayers}
\author{Yin Shi}
\email{yin.shi@iphy.ac.cn}
\affiliation{Beijing National Laboratory for Condensed Matter Physics and Institute of Physics, Chinese Academy of Sciences, Beijing 100190, China.}
\author{Mengxian Zhao}
\affiliation{Beijing National Laboratory for Condensed Matter Physics and Institute of Physics, Chinese Academy of Sciences, Beijing 100190, China.}
\affiliation{University of Chinese Academy of Sciences, Beijing 100049, China.}
\author{Fei Yang}
\affiliation{Department of Physics, The Hong Kong University of Science and Technology, Clear Water Bay, Kowloon, Hong Kong SAR.}
\author{Miao Liu}
\affiliation{Beijing National Laboratory for Condensed Matter Physics and Institute of Physics, Chinese Academy of Sciences, Beijing 100190, China.}
\author{Sheng Meng}
\email{smeng@iphy.ac.cn}
\affiliation{Beijing National Laboratory for Condensed Matter Physics and Institute of Physics, Chinese Academy of Sciences, Beijing 100190, China.}
\affiliation{University of Chinese Academy of Sciences, Beijing 100049, China.}
\affiliation{Songshan Lake Materials Laboratory, Dongguan, Guangdong 523808, China.}
\date[]{}
\begin{abstract}
Twisted cuprate bilayers have been proposed to host chiral superconductivity at temperatures comparable to their large pairing scale, yet experimental signatures of broken time-reversal symmetry remain unclear. We show that quantum phase fluctuations strongly suppress chiral order in twisted overdoped cuprate bilayers despite robust intralayer pairing. At a $45^\circ$ twist, we analytically demonstrate that the chiral transition temperature obeys $T_\chi\propto t^4$ when the single-electron interlayer tunneling amplitude $t$ is weak. This collapse originates from competition between the small Josephson energy scale supporting chirality and the temperature-dependent collective zero-point energy. Chirality consequently survives only in a narrow angular window at ultralow temperatures, with relative-phase plasma softening above the chiral transition. A distinct transport signature arising from the quantum phase fluctuations is a pronounced low-temperature enhancement of the critical current near a magic angle. These findings separate pairing and chiral ordering scales and offer a possible explanation for the difficulty of observing robust chiral superconductivity in twisted cuprates.
\end{abstract}
\maketitle

\paragraph*{Introduction.---}
\label{sec:introduction}
Chiral superconductors combine superconducting order with spontaneous breaking of time-reversal symmetry and can support topologically protected edge excitations~\cite{kallin16chiral,read00paired,PhysRevB.97.205301}. Realizing such a state at an elevated temperature would provide a route to robust topological superconductivity. Most established candidate materials, however, exhibit the relevant ordering only at very low temperatures~\cite{ishida98spin,sauls94the}. Twisted cuprate bilayers offer a different possibility: two high-temperature $d$-wave superconductors can acquire a chiral relative phase through interlayer Josephson coupling~\cite{yang18josephson,Can2021NatPhys,haenel22incoherent,volkov25josephson,pixley26twisted}.

Near a twist angle of $45^\circ$, the leading Josephson harmonic is suppressed by the sign-changing gap symmetry. Higher harmonics can then favor two degenerate relative phases of opposite chirality, producing a fully gapped $d+id'$ state. Mean-field calculations predict that this phase can extend to temperatures comparable to the superconducting transition of an isolated layer~\cite{Can2021NatPhys,tummuru22josephson}. This prediction has motivated measurements of current--phase relations, fractional Shapiro steps, and Josephson nonreciprocity in twisted cuprate junctions~\cite{zhu21presence,Zhao2023Science,martini23twisted,Wang2023NatCommun,zhu23persistent,Zhu2025NSR,confalone25cuprate}. Some observations have been interpreted as signatures of broken time-reversal symmetry, but the signatures vary across devices and thermal cycles and do not yet establish a common phase diagram. Quantitative comparisons of transition temperature, critical current density, and junction voltage are therefore essential~\cite{confalone25preserving}.

Several mechanisms modify the mean-field picture. Correlations can reduce effective tunneling~\cite{song22doping}, alternative pairing channels can compete at strong tunneling~\cite{panda26gap}, and static phase inhomogeneity can enhance higher Josephson harmonics~\cite{yuan23inhomogeneity}. These effects address the microscopic balance of pairing states. They leave open a separate fundamental question: does a large pairing amplitude ensure that the relative phase needed for chirality remains ordered? Small-cluster Hubbard-model calculations include short-wavelength correlations but cannot directly resolve the long-wavelength collective fluctuations relevant to this question~\cite{lu22doping}.

Phase fluctuations provide a natural mechanism for separating pairing and ordering scales in low-dimensional superconductors~\cite{EmeryKivelson1995,han25intrinsic,yang26microscopic,yang26preformed,PhysRevB.104.214510}. In a twisted bilayer, chirality requires a long-range-ordered, nontrivial relative phase in addition to intralayer pairing. Distinct ordering of the total and relative phases can also produce vestigial phases~\cite{liu23charge4e}. The same suppression of the leading Josephson harmonic that favors chirality also makes its phase-locking energy small. We examine this competition in overdoped cuprates using an effective continuum model~\cite{benfatto04lowenergy} and a self-consistent harmonic approximation (SCHA) for the periodic Josephson potential~\cite{feynman18statistical,fishman88role,benfatto01phase,yang26superconducting}.  
Fluctuation-induced narrowing of time-reversal-symmetry-breaking order and first-order behavior were established previously for a $t$--$J$-model monolayer~\cite{shi26quantum}. Here in a twisted nodal-bilayer Josephson junction, we find that nodal softening of the collective zero-point energy yields a chiral transition temperature proportional to tunneling's fourth power despite finite pairing.

\begin{figure*}[!t]
\centering
\includegraphics[width=\textwidth]{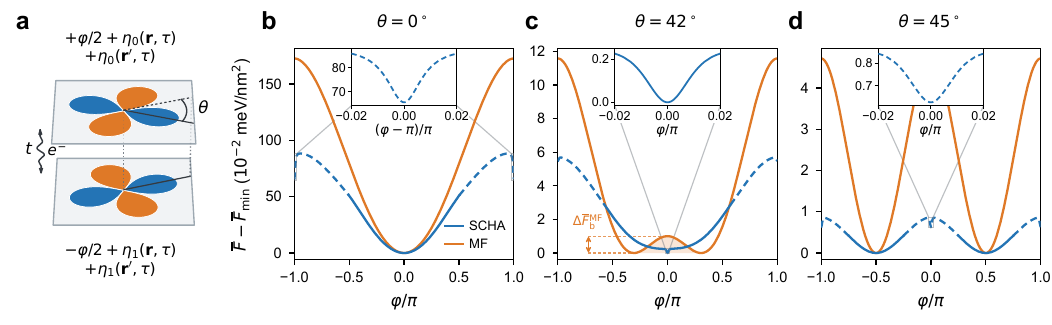}
\caption{\textbf{Quantum phase fluctuations reshape the energy landscape for chirality.}
(a) Parallel superconducting layers whose $d$-wave gap axes are relatively rotated by $\theta$. The wavy double-headed $e^{-}$ arrow represents single-electron tunneling with amplitude $t$. Opposite lobe colors indicate opposite gap signs; the pair phases are $\pm\varphi/2+\eta_l(\mathbf r,\tau)+\eta_l(\mathbf r',\tau)$, where $\eta_l$ is the fermion phase field in layer $l=0,1$; $\mathbf r$ and $\mathbf r'$ are the paired fermions' positions, and $\tau$ is imaginary time. (b--d) Globally amplitude-relaxed free energy density $\overline F(\varphi)$ at $t=10$ meV and $T/\TcL=0.245\%$ for the indicated twist angles. Blue and orange curves include and omit phase fluctuations, respectively; each curve is shifted by its own minimum $\overline F_{\min}$. In (c), pale-orange shading highlights the small mean-field barrier $\Delta{\overline F}_\mathrm{b}^\mathrm{MF}$ separating opposite-chirality minima. Solid blue segments follow a resolved positive-mass SCHA branch, whereas dashed blue segments reach the mass floor used as a numerical regulator for zero mass; the regulator is not a physical locking scale. Insets magnify the boxed SCHA dips, resolved by adaptive phase sampling and node-resolved integration. The shallow mean-field phase landscape near $45^\circ$ is especially sensitive to fluctuations.}
\label{fig:landscape}
\end{figure*}

\paragraph*{Model.---}
\label{sec:model}
Our focus is the overdoped regime, where Mott-related correlation effects are weaker than in underdoped cuprates. In this regime, low-energy transport and modeling support an effective quasiparticle description~\cite{proust02heat,lee-hone20low}. 
This motivates a low-energy continuum model without an explicit strongly correlated electron sector~\cite{Can2021NatPhys,tummuru22josephson,haenel22incoherent,volkov23current,volkov23magic,volkov25josephson}.
The system consists of two identical $d$-wave superconducting layers with relative twist angle $\theta$, coupled by incoherent single-electron tunneling~\cite{haenel22incoherent} of amplitude $t$ and long-range Coulomb interactions, as illustrated in Fig.~\ref{fig:landscape}(a). 
Their uniform gap functions are
\begin{equation}
\Delta_{0,1}(\mathbf k)=\Delta_d e^{\pm i\varphi/2}
w_{\mathbf k}\cos(2\alpha_{\mathbf k}\pm\theta),
\label{eq:gap}
\end{equation}
where $\Delta_d$ is the common pairing amplitude, $\varphi$ is the relative order-parameter phase, $\mathbf k$ is electron momentum, $\alpha_{\mathbf k}$ is its polar angle, and $w_{\mathbf k}$ restricts pairing to an energy shell around the Fermi level. A relative phase different from an integer multiple of $\pi$ breaks time-reversal symmetry and opens a gap in the otherwise nodal $d$-wave structure, with opposite signs selecting opposite chiralities.

The layer fields $\eta_l(\mathbf r,\tau)$, $l=0,1$, shown in Fig.~\ref{fig:landscape}(a) attach a dynamical phase to each fermion at in-plane position $\mathbf r$ and imaginary time $\tau$. 
A pair with fermions at $\mathbf r$ and $\mathbf r'$ therefore acquires the phase fluctuation $\eta_l(\mathbf r,\tau)+\eta_l(\mathbf r',\tau)$.
We treat these dynamical phase fluctuations using SCHA, which is most appropriate at low temperatures focused here, where smooth long-wavelength phase modes dominate, and short-ranged defects are dilute~\cite{benfatto01phase,liu23charge4e}. 
SCHA-based calculations can yield good, model-dependent estimates of transition temperatures~\cite{you12tuning,moura23theoretical,yang26selfconsistent}.
The microscopic model and local response are derived in Supplemental Material~\cite{SupplementalMaterial} Sec.~\ref{S-sec:microscopic}; the variational treatment and numerical methods are given in Supplemental Sec.~\ref{S-sec:methods}.

\paragraph*{Quantum phase fluctuations make chiral order fragile.---}
\label{sec:chiral-mechanism}
For weak tunneling, the first two Josephson harmonics expose the small energy scale that stabilizes chirality~\cite{Can2021NatPhys,tummuru22josephson,liu23charge4e}:
\begin{equation}
\begin{aligned}
    V_J(\varphi)&=J_1\cos\varphi+J_2\cos2\varphi, \\
    J_1&=O(t^2)\leq0,\qquad J_2=O(t^4)>0.
\end{aligned}
\label{eq:Vmain}
\end{equation} 
The free energy landscape becomes much shallower as the twist angle is tuned from $0$ to $45^\circ$ [Fig.~\ref{fig:landscape}(b--d)] due to the symmetry-imposed suppression of $J_1$.
At exactly $\theta=45^\circ$, $J_1=0$ and the minima of $V_J(\varphi)$ are $\varphi=\pm\pi/2$. Their mean-field advantage over $\varphi=0$ is only $2J_2$.
This quickly vanishing scale makes the chiral state susceptible to quantum phase fluctuations, much like the quantum paraelectrics in which a shallow ferroelectric potential double-well is overwhelmed by zero-point motion such that they are not ferroelectric at any finite temperatures~\cite{barrett52dielectric,muller79intrinsic,rowley14ferroelectric,yang26selfconsistent}. 
By contrast, even for an arbitrarily small positive $J_2$, mean-field theory (including thermal phase fluctuations~\cite{liu23charge4e}) would counterintuitively predict the chiral transition temperature $T_\chi$ to remain at a high energy scale, that is, the superconducting transition temperature $\TcL$ of an isolated layer~\cite{Can2021NatPhys,tummuru22josephson} (Supplemental Sec.~\ref{S-sec:phase-comparisons} and Fig.~\ref{S-fig:tunneling-comparison}). 

\begin{figure*}[!t]
\centering
\includegraphics[width=\textwidth]{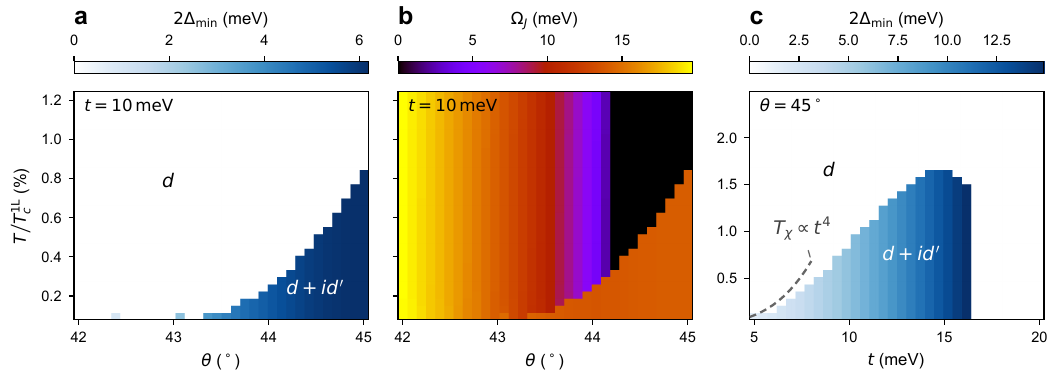}
\caption{\textbf{Chiral order survives only within a narrow, low-temperature phase-locked regime.}
(a) Minimum quasiparticle gap $2\Delta_{\min}$ in the twist-angle--temperature plane at $t=10$ meV. This map is symmetric about $45^\circ$; only one side is shown. (b) Corresponding relative-phase plasma energy $\Omega_J$. (c) Minimum gap in the tunneling--temperature plane at $\theta=45^\circ$. The gray dashed curve extrapolates the weak-tunneling law $T_\chi\propto t^4$ using the independently calculated full-model prefactor (Supplemental Sec.~\ref{S-sec:weak-tunneling}), not a fit to the displayed phase boundary. The labels $d$ and $d+id'$ classify the stable nodal and chiral variational backgrounds, respectively.}
\label{fig:phase-overview}
\end{figure*}

We focus on this $45^\circ$ twist angle to derive the weak-tunneling transition scale with quantum phase fluctuations taken into account, considering only the relative-phase sector and using the above two-harmonic Josephson model \eqref{eq:Vmain} to illustrate the mechanism (see End Matter for detailed derivation).
It follows from comparing the nodal and chiral variational free energy densities $F_d(T;t)$ and $F_\chi(T;t)$ as functions of temperature $T$ at fixed $t$. Their zero-temperature splitting is their Josephson energy difference plus a zero-point-energy contribution, giving $\Delta f_0(t)\equiv F_d(0;t)-F_\chi(0;t)=C_Qt^4+o(t^4)$, with $C_Q>0$. At temperatures well below the induced chiral quasiparticle gap $E_g\simeq t^2/\Delta_*$, where $\Delta_*$ is the finite limiting pairing amplitude, quasiparticle correction to the chiral stiffness is exponentially small but nodal quasiparticles soften the nonchiral stiffness linearly in $T$. This lowers the nonchiral zero-point energy and thus $F_d$ relative to the chiral branch by $\Gamma(t)T$ to the leading order in $T$, with a finite positive weak-tunneling limit $\Gamma(0)\equiv\lim_{t\to0}\Gamma(t)$. Equating the variational branch energies, namely demanding $\Delta f_0(t)-\Gamma(t)T=0$ gives
\begin{equation}
T_\chi\simeq\frac{\Delta f_0(t)}{\Gamma(t)}
\sim\frac{C_Q}{\Gamma(0)}t^4,
\label{eq:weakchiralTc}
\end{equation}
as $t\to0$. The pairing amplitude remains finite while $T_\chi/E_g\propto t^2\to0$, so the chiral scale collapses without collapse of pairing and the competing nodal branch can become favorable well before gapped excitations of the chiral branch are thermally populated.

Quantum phase fluctuations are decisive at this step. Although thermal quasiparticles cause the stiffness depletion, the resulting linear-in-$T$ free-energy gain comes from the quantum zero-point energy. Neither the nodal mean-field free energy nor explicit thermal occupation of the collective mode contributes a linear-in-$T$ term here. A purely static phase treatment omits this zero-point mechanism, and the small bare Josephson energy alone does not establish the transition scaling.

Figure~\ref{fig:landscape}(b--d) connects this limiting argument to finite tunneling. 
The small mean-field barrier $\Delta\overline F_{\mathrm b}^{\mathrm{MF}}$ between the opposite-chirality minima in Fig.~\ref{fig:landscape}(c) reflects a weak energy scale for chiral phase locking, despite robust intralayer pairing.
Away from exactly $\theta=45^\circ$, quantum phase fluctuations suppress the chiral-supporting second harmonic exponentially more strongly than the first [Eq.~\eqref{eq:DWmain}], allowing the chiral minima to move toward zero phase and merge [Fig.~\ref{fig:landscape}(c)]. 
At exactly $45^\circ$, chiral minima remain at the displayed low temperature but have a strongly reduced locking scale. 
At higher temperatures symmetric sampling of opposite chiralities can restore time-reversal symmetry without selecting a coherent zero-relative-phase state. 

\paragraph*{Chirality survives only at ultralow temperatures.---}
\label{sec:phase-diagram}
Now we obtain eight Josephson harmonics by Fourier projection of the unexpanded microscopic free energy, take into account both the relative-phase and overall-phase fluctuations, and determine the amplitude, variance, and positive trial mass self-consistently. 
We use this calculation to examine the phase diagram, collective locking scale, and Josephson transport.

Figure~\ref{fig:phase-overview} brings together the fluctuation-corrected phase diagram and its collective locking scale. We identify the chiral background through the minimum quasiparticle excitation gap $2\Delta_{\min}$, which distinguishes fully gapped $d+id'$ order from the nodal $d$-wave state. 
Temperatures are expressed as a percentage of the monolayer mean-field transition temperature $\TcL$, a pairing reference scale rather than a calculated vortex-unbinding temperature.

At realistic $t=10$ meV~\cite{tummuru22josephson}, chirality survives only within approximately $2^\circ$ of $\theta=45^\circ$ and below $0.8\%$ of $\TcL$ [Fig.~\ref{fig:phase-overview}(a)]. A large intralayer pairing scale therefore does not protect the chiral relative phase over a comparable temperature interval. The abrupt gap change at the phase boundary is a first-order phase-transition behavior arising from competition between distinct nodal and chiral free-energy minima, whose crossing produces a discontinuous change in the preferred state. Whether fluctuations beyond the retained approximation modify this transition remains an open question. Detailed comparisons with mean-field theory and the scans at other tunneling amplitudes are given in Supplemental Sec.~\ref{S-sec:phase-comparisons} and Figs.~\ref{S-fig:phase-comparison} and \ref{S-fig:tunneling-comparison}.

The relative-phase mode distinguishes weak locking from disappearance of pairing. 
It can acquire a gap $\Omega_J$ [Fig.~\ref{fig:phase-overview}(b)], called the Josephson plasma frequency, which is increased by the electrostatic cost of transferring charge between layers.
Its spectroscopic visibility also depends on electromagnetic dynamics and geometry beyond the present instantaneous-Coulomb model~\cite{gabriele24charge,sellati25ghost}, as discussed in Supplemental Sec.~\ref{S-sec:smallq-dispersion}.
Comparing Fig.~\ref{fig:phase-overview}(a) and (b) shows that strong $\Omega_J$ softening occurs above the chiral boundary and within a very narrow angular range of approximately $0.8^\circ$ around $\theta=45^\circ$, while the pairing amplitude varies much less (Supplemental Fig.~\ref{S-fig:stiffness}). This softening is consistent with relative-phase unlocking due to the extremely shallow Josephson potential [Fig.~\ref{fig:landscape}(c)] overwhelmed by phase fluctuations, rather than a rigid state with $\varphi=0$. The soft mode therefore separates loss of interlayer locking from loss of the pairing amplitude. Supplemental Fig.~\ref{S-fig:plasma-comparison} compares the SCHA and mean-field plasma gaps at the same tunneling amplitude.

Increasing tunneling does not restore a high-temperature chiral state [Fig.~\ref{fig:phase-overview}(c)]. At $\theta=45^\circ$, the chiral transition temperature follows the $t^4$ law [Eq.~\eqref{eq:weakchiralTc}] at small $t$'s and then varies nonmonotonically with $t$ and remains below approximately $1.6\%$ of $\TcL$ throughout the displayed scan. The turnover is consistent with a competition in the low-temperature ratio $T_\chi\simeq\Delta f_0(t)/\Gamma(t)$: stronger tunneling enhances Josephson stabilization ($\Delta f_0$) but also flattens the competing nonchiral state's nodes, increasing its temperature-induced stiffness depletion and lowering its collective zero-point energy ($\Gamma$) more rapidly. Amplitude relaxation reinforces this latter effect, so $T_\chi$ can decrease even while $J_2$ grows (Supplemental Sec.~\ref{S-sec:tunneling-comparison}). 
Together, the three panels identify a narrow regime where chirality survives despite a much larger pairing scale.

\paragraph*{Josephson critical current.---}
\label{sec:critical-current}
\begin{figure}[tbp]
\centering
\includegraphics[width=\columnwidth]{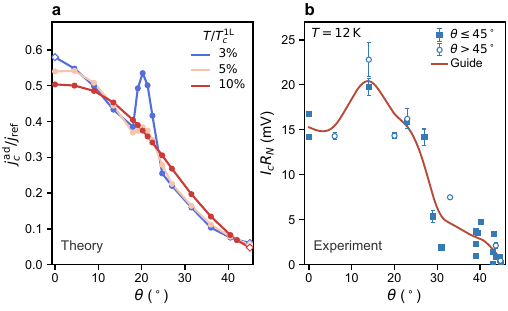}
\caption{\textbf{Angular Josephson-current enhancement and thermal smearing.}
(a) Adiabatic current-biased limits at $t=10$ meV and $T/\TcL=3\%,5\%,10\%$, including downhill recapture and normalized by $j_{\mathrm{ref}}=j_c^{\mathrm{MF}}(T=0,\theta=0)$. Filled circles denote positive-mass terminal states; open diamonds denote provisional floor-regulated estimates. Lines connect the 16 sampled angles. (b) Original 12 K data from Zhao et al.~\cite{Zhao2023Science} for twisted Bi$_2$Sr$_2$CaCu$_2$O$_{8+x}$ flakes, folded about $45^\circ$. Filled squares and open circles distinguish angles below and above $45^\circ$ before folding. Vertical bars follow the original data-processing procedure; the spline only guides the eye. Definitions, processing, and stability checks are given in Supplemental Sec.~\ref{S-sec:current-bias}.}
\label{fig:current}
\end{figure}

The current-biased junction poses a nonequilibrium stability problem, distinct from the zero-current equilibrium phase diagram. We estimate its adiabatic static-current limit: for an imposed current density $j_b$, static states are local minima of the tilted variational free energy density
\begin{equation}
\begin{aligned}
G_{j_b}&=F_{\mathrm{SCHA}}-\frac{\hbar}{2e}j_b\varphi,\\
j(\varphi)&=\frac{2e}{\hbar}\frac{d\overline F}{d\varphi}=j_b,
\end{aligned}
\label{eq:current}
\end{equation}
Here $F_{\mathrm{SCHA}}$ is the variational free energy density, $e>0$ the elementary charge, and $\overline F$ its internally relaxed branch free energy at fixed $T,t,\theta$. Starting from the lowest-energy stable zero-current state, we increase $j_b$ and allow downhill recapture after an instability; $j_c^{\mathrm{ad}}$ is the final reachable static limit (see Supplemental Sec.~\ref{S-sec:current-bias} for numerical details). 

Figure~\ref{fig:current}(a) shows an intermediate-angle current ramp near $\theta\simeq20^\circ$ at $T/\TcL=3\%$. This enhancement occurs near a tunneling-shifted merger of quasiparticle nodes of the two layers, which defines the magic angle~\cite{volkov23magic}. The merger flattens the low-energy dispersion and enhances the density of states, making the response to imposed-phase-induced quasiparticle gapping unusually strong~\cite{volkov23current}. The rapid recovery of stiffness and increase of overall-phase zero-point energy, together with amplitude relaxation, steepen the full variational phase landscape and enhance the current (Supplemental Sec.~\ref{S-sec:current-node-merger}).
The narrow dips in Fig.~\ref{fig:landscape}(b--d) illustrate such steep local phase dependence at $T/\TcL=0.245\%$ and different twist angles. Thermal smearing reduces the ramp to a weak hump at $T/\TcL=5\%$ and removes it on the sampled grid at $T/\TcL=10\%$. The location of the node merger and associated current ramp is parameter dependent.

Since caused by quasiparticle gapping, this intermediate-angle current ramp can serve as a clear transport signature of current-induced topological superconductivity in twisted nodal superconductors.
Some intermediate-angle experimental data on twisted cuprate Josepshon junctions~\cite{Wang2023NatCommun,Zhao2023Science}, including Zhao et al.'s symmetry-folded 12 K data [Fig.~\ref{fig:current}(b)], show enhancements qualitatively reminiscent of this ramp. Here the measured quantity is $I_cR_N$, the product of junction critical current $I_c$ and normal resistance $R_N$, not current density. Sample variations, finite flake thickness, and differing temperatures and junction parameters preclude identifying this resemblance as confirmation. Systematic low-temperature angular scans could test its thermal evolution.

\paragraph*{Conclusion.---}
\label{sec:conclusion}
Quantum phase fluctuations strongly suppress chiral superconductivity in twisted overdoped cuprate bilayers despite a robust intralayer pairing amplitude. Near $45^\circ$, chirality relies on a weak second Josephson harmonic that is exponentially reduced by phase fluctuations, and zero-point energy of phase fluctuations subsequently make the chiral transition scale vanishes as $T_\chi\propto t^4$ in the weak-tunneling limit. At realistic $t=10$ meV, chirality survives within approximately $2^\circ$ of $45^\circ$ and below $0.8\%$ of $\TcL$, with relative-phase plasma softening above the chiral transition. Pairing and chiral locking therefore have widely separated scales. Furthermore, the angular critical current develops a low-temperature intermediate-angle ramp that is washed out on warming.
This separation offers a possible reconciliation of conflicting experimental signatures with mean-field predictions of high-temperature chirality: fluctuation-suppressed chiral ordering may occur below the temperatures reached in some existing measurements, despite robust pairing.

Combined phase-sensitive, plasma, and critical-current measurements can test this separation of scales and current-induced topology. 
Robust chirality requires interlayer phase coherence, not merely strong pairing.


\begin{acknowledgments}
This work was supported by the start-up grant from the Institute of Physics, Chinese Academy of Sciences, and CAS Project for Young Scientists in Basic Research (Grant No. YSBR143). S.M. acknowledges support from the Ministry of Science and Technology (Grant No. 2021YFA1400200), the National Natural Science Fund of China (Grant Nos. 12450401 and 12025407), and the Chinese Academy of Sciences (No. YSBR047).
\end{acknowledgments}

%

\clearpage
\section*{End Matter}
\label{sec:end-matter}
\appendix*
\setcounter{equation}{0}
\renewcommand{\theequation}{A\arabic{equation}}
\renewcommand{\theHequation}{A\arabic{equation}}
\paragraph*{Weak-tunneling transition scale.---}
The relative-phase fluctuation is $\delta\varphi(\mathbf r, \tau)=2[\eta_0(\mathbf r, \tau)-\eta_1(\mathbf r, \tau)]$. In units $k_B=\hbar=1$, its minimal quantum action is
\begin{equation}
\begin{aligned}
S_-={}&\frac{T}{2}\sum_n\int_{\mathbf q}
\left[\chi_-(\mathbf q)\nu_n^2+\rho_-|\mathbf q|^2\right]
|\delta\varphi(\mathbf q,\nu_n)|^2 \\
&+\int_0^\beta d\tau\int d^2r\,
V_J(\varphi+\delta\varphi).
\end{aligned}\label{eq:intuitivequantumaction}
\end{equation}
Here $T$ is temperature, $\beta=1/T$, and $\nu_n=2\pi nT$ is a bosonic Matsubara frequency with integer $n$; $\delta\varphi(\mathbf q,\nu_n)$ is the Fourier transform of the fluctuation field. We write $\int_{\mathbf q}=\int_{|\mathbf q|<\Lambda}d^2\mathbf q/(2\pi)^2$, with momentum cutoff $\Lambda=\Delta_d/v_F$ set by the inverse coherence-length scale, where $v_F$ is the Fermi velocity. The minus subscript denotes the relative-phase sector: $\rho_-$ is its spatial stiffness and $\chi_-(\mathbf q)>0$ its temporal susceptibility after relative-density fluctuations and Coulomb interactions have been integrated out. These coefficients are obtained from the fermions, retaining their dependence on temperature, pairing amplitude, and background phase. The temporal term retains quantum dynamics. 
We use a common fixed pairing amplitude for the two backgrounds.

We treat the periodic potential using SCHA. Introduce a positive trial Josephson mass $r_J$ by replacing the nonlinear Josephson potential with $r_J(\delta\varphi)^2/2$. Semicolons separate variational or evaluation arguments from parameters held fixed in the calculation; unshown dependencies are implicit. The trial mode energy $\omega_-$ and equal-time phase variance $W=\langle(\delta\varphi)^2\rangle_G$ are
\begin{align}
\omega_-(\mathbf q)&=\sqrt{\frac{\rho_-|\mathbf q|^2+r_J}{\chi_-(\mathbf q)}},\\
W(r_J;T,t)&=\int_{\mathbf q}\frac{1}{2\chi_-(\mathbf q)\omega_-(\mathbf q)}
\coth\!\left[\frac{\omega_-(\mathbf q)}{2T}\right],
\label{eq:intuitivetrialmode}
\end{align}
where $\langle\cdot\rangle_G$ denotes averaging in this Gaussian trial state. At fixed amplitude and cutoff, the $t$ dependence enters through the microscopic response coefficients.
The collective-mode gap is $\Omega_J\equiv\omega_-(\mathbf 0)=[r_J/\chi_-(\mathbf 0)]^{1/2}$. 
The trial free energy density is
\begin{equation}
F_G(r_J;T,t)=\int_{\mathbf q}\left[
\frac{\omega_-(\mathbf q)}{2}+T\ln(1-e^{-\omega_-(\mathbf q)/T})\right].
\label{eq:intuitiveGaussianfreeenergy}
\end{equation}
For background $s=d,\chi$, the zero-point contribution at any temperature is
\begin{equation}
E_{G,s}(r_J;T,t)=\frac12\int_{\mathbf q}
\omega_{-,s}(\mathbf q,r_J;T,t).
\label{eq:temperaturedependentEG}
\end{equation}
It depends on $T$ through the fermionic response coefficients $\rho_{-,s}(T;t)$ and $\chi_{-,s}(\mathbf q;T,t)$. Zero-point refers to oscillator occupation, not to evaluating these coefficients at $T=0$. Only with the coefficients, cutoff, and trial mass frozen would this term be temperature independent. Thus $F_G$ contains $E_G(r_J;T,t)$, not a fixed $T=0$ zero-point energy, in addition to the explicit Bose occupation term.

Writing $F_{\mathrm{av}}(T;t)$ for the phase-independent Fourier component of the mean-field free energy at temperature $T$, the variational free energy density is~\cite{feynman18statistical,fishman88role,benfatto01phase}
\begin{equation}
\begin{aligned}
F_{\mathrm{var}}(\varphi,r_J;T,t)={}&F_{\mathrm{av}}(T;t)+F_G(r_J;T,t) \\
&-\frac{r_JW}{2}+\langle V_J(\varphi+\delta\varphi;T,t)\rangle_G.
\end{aligned}\label{eq:intuitiveFvar}
\end{equation}
Here $W=W(r_J;T,t)$; $T$ and $t$ are held fixed during the variational optimization.
For an integer harmonic index $n$, Gaussian averaging gives
\begin{equation}
\langle\cos[n(\varphi+\delta\varphi)]\rangle_G
=\cos(n\varphi)e^{-n^2W/2}.
\label{eq:DWmain}
\end{equation}
Thus the second harmonic is reduced by the Debye--Waller factor $Z=e^{-2W}$. At $\theta=45^\circ$, its local stabilization energy density $\delta f_\chi$ and stationary chiral trial mass obey
\begin{equation}
\delta f_\chi=2J_2Z,\qquad r_J=4J_2Z,
\label{eq:chiralscale}
\end{equation}
which are exponentially suppressed by the phase fluctuations.

We first determine the zero-temperature free-energy splitting between the nodal and chiral phases. Let $\Delta_*$ and $\rho_*$ be the finite limiting pairing amplitude and relative-phase stiffness as $t\to0$. In this limit the quantum variance at zero trial mass is
\begin{equation}
W_* =\int_{\mathbf q}\frac{1}{2|\mathbf q|\sqrt{\rho_*\chi_-(\mathbf q)}}<\infty,
\qquad Z_*=e^{-2W_*},
\label{eq:intuitivequantumvariance}
\end{equation}
where $\chi_-$ and $\Lambda$ are evaluated at zero temperature and the limiting amplitude. 
The integral does not have infrared divergence because $\chi_-(\mathbf q)$ has a finite positive limit at $\mathbf q=\mathbf 0$ in two dimensions. 
Hence $Z_*$ is finite and a small positive chiral mass can survive for arbitrarily weak tunneling. Write $J_2=b_4t^4+o(t^4)$, with finite positive coefficient $b_4$ derived in Supplemental Sec.~\ref{S-sec:weak-tunneling}. Equation~\eqref{eq:chiralscale} then gives $r_J\sim4b_4Z_*t^4$, while the mode gap scales as $\Omega_J\propto t^2$. Quantum suppression of the local harmonic alone therefore does not imply a nonzero-tunneling quantum critical point.

After trial-mass selection, $T$ becomes the evaluation argument of each branch free energy. At fixed tunneling amplitude $t$, let $F_d(T;t)$ and $F_\chi(T;t)$ denote the variational free energy densities at temperature $T$, centered on the nodal background $\varphi=0$ and chiral background $\varphi=\pi/2$, respectively.
At $T=0$, $F_G(r_J;0,t)=E_G(r_J;0,t)$. For either background $s=d,\chi$, let $W_s$ be its phase variance. At $T=0$ and fixed microscopic response coefficients, $\partial_{r_J}E_{G,s}(r_J;0,t)=W_s(r_J;0,t)/2$. Since $W_s(r_J;0,t)\to W_*$ in the joint weak-tunneling, small-mass limit, the combination entering $F_{\mathrm{var}}$ satisfies
\begin{equation}
E_{G,s}(r_J;0,t)-\frac{r_JW_s(r_J;0,t)}{2}
=E_{G,s}(0;0,t)+o(r_J).
\label{eq:zeropointmasscancellation}
\end{equation}
Only the leading mass-induced shift cancels, not the zero-point background. To obtain its leading phase-dependent splitting, compare the two backgrounds at a common amplitude tending to $\Delta_*$ and a common cutoff. For the isotropic local response retained here, angular integration gives their zero-mass energies as
\begin{equation}
E_{G,s}(0;0,t)=\frac{1}{4\pi}\int_0^\Lambda d|\mathbf q|\,|\mathbf q|^2
\sqrt{\frac{\rho_{-,s}(0;t)}{\chi_{-,s}(\mathbf q;0,t)}}.
\label{eq:masslessEGbackground}
\end{equation}
The retained local kernels have common order-$t^2$ corrections, but their order-$t^2$ phase-dependent terms are first Josephson harmonics and vanish after angular integration at $\theta=45^\circ$. The second Josephson harmonic first appears at fourth order in tunneling and distinguishes $\varphi=0$ from $\pi/2$. Using the finite response limits established in Supplemental Sec.~\ref{S-sec:weak-tunneling}, write
\begin{align}
\delta\rho_-(t)&\equiv\rho_{-,d}(0;t)-\rho_{-,\chi}(0;t)
\nonumber\\
&=\rho_{-,4}t^4+o(t^4),
\\
\delta\chi_-(\mathbf q;t)&\equiv\chi_{-,d}(\mathbf q;0,t)-\chi_{-,\chi}(\mathbf q;0,t)
\nonumber\\
&=\chi_{-,4}(\mathbf q)t^4+o(t^4).
\label{eq:relativeResponseSplitting}
\end{align}
Let $\chi_{-,*}(\mathbf q)$ denote the common $t\to0$ screened susceptibility. Expanding the square root in Eq.~\eqref{eq:masslessEGbackground} to first order in these differences gives
\begin{align}
&E_{G,d}(0;0,t)-E_{G,\chi}(0;0,t)
=g_4t^4+o(t^4),
\label{eq:EGbackgroundSplitting} \\
&g_4=\frac{1}{8\pi}\int_0^\Lambda d|\mathbf q|\,|\mathbf q|^2
\left[\frac{\rho_{-,4}}{\sqrt{\rho_*\chi_{-,*}(\mathbf q)}}-\frac{\sqrt{\rho_*}\,\chi_{-,4}(\mathbf q)}{\chi_{-,*}^{3/2}(\mathbf q)}\right].
\label{eq:g4relative}
\end{align}
The cutoff in $g_4$ is evaluated at the limiting amplitude. This scaling uses finite, node-resolved response coefficients; symmetry alone does not guarantee those limits or fix the sign of $g_4$. For the parameters studied here, node-resolved numerical tests give $g_4>0$, as detailed in Supplemental Sec.~\ref{S-sec:weak-tunneling}. Adding the averaged Josephson energy contribution yields the zero-temperature chiral advantage
\begin{equation}
\begin{aligned}
\Delta f_0(t)&\equiv F_d(0;t)-F_\chi(0;t)=C_Qt^4+o(t^4),
\\
C_Q&=g_4+2b_4Z_*>0.
\end{aligned}\label{eq:CQmain}
\end{equation}

Next we determine how temperature removes this advantage. The chiral quasiparticles have a small induced gap $E_g\simeq t^2/\Delta_*$, whereas the nonchiral background remains nodal. For $T\ll E_g$, quasiparticle corrections in the chiral state are exponentially small, but nodal quasiparticles soften the nonchiral stiffness linearly in $T$~\cite{benfatto01phase}. In our continuum model, with chemical potential $\mu$,
\begin{equation}
\rho_{-,d}(T)=\rho_{-,d}(0)-a_-T+o(T),
\qquad a_-=\frac{\mu\ln(2)}{4\pi\Delta_*},
\label{eq:nodalstiffnessintuition}
\end{equation}
while $\chi_-(\mathbf q)$ has no linear-in-$T$ term. Inserting this softening into Eq.~\eqref{eq:temperaturedependentEG}, at fixed trial mass, amplitude, and cutoff, gives $E_{G,d}(r_J;T,t)-E_{G,d}(r_J;0,t)=-\Gamma(t)T+o(T)$. Thus the zero-point term itself has a linear temperature dependence, with positive coefficient
\begin{equation}
\Gamma(t)
=a_-\left.\frac{\partial E_{G,d}}{\partial\rho_-}\right|_{\chi_-,\,r_J,\,\Lambda}
=\frac{a_-}{4}\int_{\mathbf q}\frac{|\mathbf q|^2}{\chi_-(\mathbf q)\omega_-(\mathbf q)}>0,
\label{eq:Gammasteps}
\end{equation}
All quantities in this derivative are evaluated at $T=0$ on the nodal background. 
Variational stationarity removes the implicit temperature derivative of the trial mass from this free-energy slope.
$\Gamma(0)$ is still finite and positive.

The competition is therefore
\begin{equation}
F_d-F_\chi=\Delta f_0(t)-\Gamma(t)T+o(T).
\label{eq:lowTcompetition}
\end{equation}
Define $T_\chi$ as the temperature where these prescribed variational branches cross. Equating their free energies gives main-text Eq.~\eqref{eq:weakchiralTc} as $t\to0$.
The chiral trial mass and plasma gap can remain finite at this crossing, which need not coincide with loss of the locked solution.

\arxivStartSupplement
\arxivSupplementTitle{Supplemental Material: Quantum phase fluctuations quench chiral superconductivity in twisted overdoped cuprate bilayers}{Yin Shi, Mengxian Zhao, Fei Yang, Miao Liu, and Sheng Meng}

\noindent This document supplies the microscopic derivation and supporting checks for the main article. References use the numbering of the single References and Notes list in the main manuscript. Equations without an S prefix are in the Letter or its End Matter. 

\arxivSupplementContents

\section{Microscopic model and local response}
\label{S-sec:microscopic}
\subsection{Continuum model and parameters}
\label{S-sec:continuum-model}
Our focus is the overdoped regime, where Mott-related correlation effects are weaker than in underdoped cuprates and low-energy transport and modeling support an effective quasiparticle description~\cite{proust02heat,lee-hone20low}. This motivates a continuum model without an explicit strongly correlated electron sector. Residual electronic renormalizations are incorporated phenomenologically in the band, tunneling, and pairing parameters, while long-range Coulomb interactions and collective phase fluctuations are retained. The approximation does not require electronic correlations to vanish.

We work with a continuum microscopic model of coupled $d$-wave superconducting monolayers, which has been checked against the lattice model at the mean-field level~\cite{Can2021NatPhys,tummuru22josephson}.
We use a convenient notation $r\equiv (i\tau, \mathbf{r})$, where $\tau$ is the imaginary time and $\mathbf{r}$ is the in-plane coordinate.
Its reciprocal counterpart is $k \equiv (i\omega, \mathbf{k})$ (for fermions) or $q \equiv (i\nu, \mathbf{q})$ (for bosons).
The action in real space reads

\begin{equation}
\begin{aligned}
    S ={}& \sum_{l\sigma} \int dr \psi^\dagger_{l\sigma}(r) \biggl( \partial_\tau - \frac{\hbar^2\nabla^2}{2m} - \mu \biggr) \psi_{l\sigma}(r) - \sum_{ll'\sigma} \iint dr dr' \delta(\tau-\tau') t_{l-l'}(\mathbf{r}-\mathbf{r}') \psi^\dagger_{l\sigma}(r) \psi_{l'\sigma}(r') \\
    &- \sum_l\iint dr dr' \delta(\tau-\tau') g_l(\mathbf{r}, \mathbf{r}') \psi^\dagger_{l\uparrow} (r) \psi^\dagger_{l\downarrow} (r') \psi_{l\downarrow}(r') \psi_{l\uparrow}(r) + \frac{1}{2}\sum_{ll'}\iint dr dr' \delta(\tau-\tau') V_{l-l'}(\mathbf{r}-\mathbf{r}') n_l(r) n_{l'}(r').
\end{aligned}
\end{equation}

Here $\psi_{l\sigma}(r)$ is the electron annihilation operator, where $l=0,1$ is the layer index and $\sigma$ is the spin index. 
$m$ is the effective mass and $\mu$ is the chemical potential, with band dispersion $\xi_{\mathbf k}=\hbar^2|\mathbf k|^2/(2m)-\mu$.
$t_l(\mathbf{r})$ is the interlayer tunneling amplitude, approximated as $t_l(\mathbf{r})=t_l\delta(\mathbf{r})$ with $t_0=0$ and $t_1=t$. This local-tunneling approximation neglects an additional orbital form factor~\cite{Can2021NatPhys,tummuru22josephson,song22doping,haenel22incoherent}.
The intralayer $d$-wave pairing interaction is separable:
\begin{equation}
g_l(\mathbf k,\mathbf k')=\frac{g}{\Omega}
c_l(\mathbf k)c_l(\mathbf k'),
\qquad c_l(\mathbf k)=w_{\mathbf k}\cos(2\alpha_{\mathbf k}+\theta_l),
\label{S-eq:pairing}
\end{equation}
Here $\alpha_{\mathbf k}$ is the polar angle, $\theta_{0,1}=\pm\theta$, $g>0$ is the pairing strength, and $\Omega$ is the junction interface area. We use the sharp pairing window $w_{\mathbf k}=\Theta(\varepsilon_c-|\xi_{\mathbf k}|)$. The long-range Coulomb potentials are
\begin{equation}
V_0(\mathbf q)=\frac{e^2}{2\epsilon_0\epsilon_b |\mathbf q|},
\qquad V_1(\mathbf q)=V_0(\mathbf q)e^{-|\mathbf q|d},
\label{S-eq:Coulomb}
\end{equation}
with $\widetilde V_{0,\pi}=V_0\pm V_1$, layer spacing $d$, and background dielectric constant $\epsilon_b$~\cite{shao24electrical}.

$n_l(r)=\sum_{\sigma}\psi_{l\sigma}^\dagger(r)\psi_{l\sigma}(r)$ is the electron density.

We set $k_B=1$ in analytical expressions. The physical parameters are $m=5m_e$, $\mu=196$ meV, $\varepsilon_c=60$ meV, $\epsilon_b=4.5$, $d=12.8$ \AA, and $a=5.4$ \AA. The pairing strength $g/a^2=480$ meV gives a monolayer zero-temperature mean-field amplitude of approximately 40 meV~\cite{Can2021NatPhys}.

\subsection{Phase fields and the fermionic effective action}
\label{S-sec:phase-action}
We extend the phase-action construction previously used for a monolayer $t$--$J$ model~\cite{shi26quantum}.
Using the Hubbard-Stratonovich transformation, we introduce auxiliary fields $\Delta_l(r,r')$ and $\phi_l(r)$ to decouple the pairing interaction and Coulomb repulsion, respectively.
Since the gapless phase fluctuations dominate, the superconducting gap field can be decoupled into a stationary part and a dynamical phase, $\Delta_l(r,r')\approx\Delta_l(\mathbf{r}-\mathbf{r}')\operatorname{e}^{i[\eta_l(r)+\eta_l(r')]}$~\cite{paramekanti00effective,shi26quantum}.
Next, we switch to the standard Nambu-spinor representation for each layer $\Psi_l(r)=(\psi_{l\uparrow}(r), \psi^\dagger_{l\downarrow}(r))^{\operatorname{T}}$, perform a local gauge transformation $\Psi_l(r)\rightarrow \operatorname{e}^{i\eta_l(r)\sigma_3}\Psi_l(r)$ to eliminate the dynamical phase of $\Delta_l(r,r')$, and work in full Fourier space 
\begin{align}
    \tilde\Psi_\kappa(k) &= \sqrt{\frac{T}{\Omega N_z}}\sum_{l}\int dr \operatorname{e}^{-i(k\cdot r+\kappa l)}\Psi_l(r), \\
    \tilde\eta_\kappa(q) &= \sqrt{\frac{T}{\Omega N_z}}\sum_{l}\int dr \operatorname{e}^{-i(q\cdot r+\kappa l)}\eta_l(r), \\
    \tilde\Delta_\kappa(\mathbf{k}) &= N_z^{-1}\sum_l\int dr \operatorname{e}^{-i(\mathbf{k}\cdot\mathbf{r}+\kappa l)}\Delta_l(\mathbf{r})= N_z^{-1}\sum_l\operatorname{e}^{-i\kappa l} \Delta_l(\mathbf{k}), \\
    \Delta_l(\mathbf{k}) &= \Delta_l w_\mathbf{k} \cos(2\alpha_\mathbf{k}+\theta_l), \\
    \tilde\phi_\kappa(q) &= \sum_l\int dr \operatorname{e}^{-i(q\cdot r+\kappa l)}\phi_l(r), \\
    \tilde V_\kappa(\mathbf{q}) &= \sum_l \operatorname{e}^{-i\kappa l}V_l(\mathbf{q}), \\
    \tilde{t}_\kappa &= \sum_l \operatorname{e}^{-i\kappa l} t_l.
\end{align}
Here $\kappa=0,\pi$. 
$\Delta_l$ are the order parameters, which are potentially complex numbers.
$N_z=2$ is the number of layers.
$\sigma_1,\sigma_2,\sigma_3$ are Pauli matrices and $T$ is the temperature in energy units.
The displayed Green-function blocks use the convention $\sigma_2=\left(\begin{smallmatrix}0&i\\-i&0\end{smallmatrix}\right)$, consistent with the original layer--Nambu derivation.
Throughout the paper, a tilde over a symbol means its Fourier component in full Fourier space.

The resultant action is

\begin{equation}
    S = -\tilde{\Psi}^\dagger(G^{-1}-\Sigma)\tilde{\Psi} + \frac{T}{2\Omega N_z}\sum_{\kappa q}\frac{\tilde{\phi}_\kappa(q)\tilde{\phi}_{-\kappa}(-q)}{\tilde{V}_\kappa(\mathbf{q})}  + \frac{\Omega}{Tg}\sum_l|\Delta_l|^2, 
\label{S-eq:S}
\end{equation}
where $\tilde{\Psi}$ represents a vector in the direct product of the Nambu, layer-index, and $k$ spaces.
The (inverted) mean-field Green's function, including \emph{all} orders of Josephson coupling, is
\begin{equation}
    G^{-1}_{\kappa\kappa'}(k,k') = \delta_{kk'}\{\delta_{\kappa\kappa'}[ i\omega - (\xi_\mathbf{k} - \tilde{t}_\kappa )\sigma_3 ] - \operatorname{Re}\tilde{\Delta}_{\kappa-\kappa'}(\mathbf{k})\sigma_1 + \operatorname{Im}\tilde{\Delta}_{\kappa-\kappa'}(\mathbf{k}) \sigma_2 \}\equiv \delta_{kk'}G^{-1}_{\kappa\kappa'}(k), \label{S-eq:G}
\end{equation}
$\Sigma=\Sigma'+\Sigma''$ is the self-energy arising from phase fluctuations, where $\Sigma'$ contains the gradient and Coulomb terms,
\begin{equation}
\begin{aligned}
    \Sigma'_{\kappa\kappa'}(k,k') ={}& \sqrt{\frac{T}{\Omega N_z}}\tilde{\eta}_{\kappa-\kappa'}(k-k')\biggl[ (\omega-\omega')\sigma_3 + \frac{i\hbar^2\mathbf{k}'\cdot(\mathbf{k}-\mathbf{k}')}{m} \biggr] + \frac{T}{\Omega N_z}i\tilde{\phi}_{\kappa-\kappa'}(k-k')\sigma_3 \\
    &+ \frac{T}{2\Omega N_z} \sum_{\kappa''k''} \frac{\hbar^2}{m}(\mathbf{k}''-\mathbf{k})\cdot(\mathbf{k}''-\mathbf{k}') \tilde{\eta}_{\kappa-\kappa''}(k-k'')\tilde{\eta}_{\kappa''-\kappa'}(k''-k')\sigma_3,
\end{aligned}
\label{S-eq:Sigma}
\end{equation}
and $\Sigma''$ contains Josephson phase fluctuations,
\begin{equation}
    \Sigma''_{\kappa\kappa'}(k-k') = \frac{-T}{\Omega N_z} \int dr \operatorname{e}^{-i(k-k')\cdot r} t \bigl[ \operatorname{e}^{i\kappa'} (\operatorname{e}^{-i\sqrt{N_z}\eta_\pi(r)\sigma_3}-1) + \operatorname{e}^{-i\kappa} (\operatorname{e}^{i\sqrt{N_z}\eta_\pi(r)\sigma_3}-1) \bigr]\sigma_3,
\end{equation}
which depends only on the relative phase between the two layers $\eta_\pi(r)=[\eta_0(r)-\eta_1(r)]/\sqrt{N_z}$.

Integrating out $\tilde{\Psi}$ in the action~(\ref{S-eq:S}), we have
\begin{equation}
    S = -\operatorname{Tr}[\ln(G^{-1}-\Sigma)] + \frac{T}{2\Omega N_z}\sum_{\kappa k}\frac{\tilde{\phi}_\kappa(k)\tilde{\phi}_{-\kappa}(-k)}{\tilde{V}_\kappa(\mathbf{k})} + \frac{\Omega}{Tg}\sum_l|\Delta_l|^2.
\end{equation}
Expanded in powers of $G\Sigma$, $S$ has three contributions $S = F_0\Omega/T + S'_f + S''_f$.
Here 
\begin{equation}
    F_0(\Delta_d,\varphi;T,t,\theta) = -\frac{T}{\Omega}\operatorname{Tr}[\ln(G^{-1})] + \frac{1}{g}\sum_l|\Delta_l|^2= \frac{-1}{\Omega}\sum_{\kappa\mathbf{k}} [2T\ln(1+e^{-E_\kappa(\mathbf{k})/T})+E_\kappa(\mathbf{k})] + \sum_l\frac{|\Delta_l|^2}{g}, \label{S-eq:FMF}
\end{equation}
is the mean-field free energy density, where $\pm E_\kappa(\mathbf{k})$ are the poles of $G_{\kappa\kappa'}(k)$,
\begin{equation}
    E_{0(\pi)}^2(\mathbf{k}) = \frac{|\Delta_0(\mathbf{k})|^2+|\Delta_1(\mathbf{k})|^2}{2}+\xi_\mathbf{k}^2+t^2\pm \sqrt{\frac{(|\Delta_0(\mathbf{k})|^2-|\Delta_1(\mathbf{k})|^2)^2}{4}+t^2(|\Delta_0(\mathbf{k})-\Delta_1(\mathbf{k})|^2+4\xi_\mathbf{k}^2)}.\label{S-eq:bands}
\end{equation}
\(S'_f\) describes phase-gradient fluctuations, which we treat within the local gradient--gradient approximation by neglecting mixed gradient--tunneling and nonlocal tunneling contributions~\cite{sun20collective,sellati23generalized,yang26superconducting}, justified by the weak interlayer tunneling.
\(S''_f\) describes the local relative-phase fluctuations.
We stress that we are interested only in the low-temperature regime, where only low-energy/long-wavelength fluctuations are relevant~\cite{paramekanti00effective,shi26quantum}.
Therefore, it is valid to retain terms up to the quadratic order of $q\tilde{\eta}_\kappa(q)$ and $\tilde{\phi}_\kappa(q)$ in $S'_f$,
\begin{align}
    S'_f &\approx \operatorname{Tr}[G\Sigma']+\frac{1}{2}\operatorname{Tr}[(G\Sigma')^2]+\frac{T}{2\Omega N_z}\sum_{\kappa k}\frac{\tilde{\phi}_\kappa(k)\tilde{\phi}_{-\kappa}(-k)}{\tilde{V}_\kappa(\mathbf{k})}, \nonumber \\
    &\approx \frac{1}{2}\sum_{\kappa\kappa' q}\biggl[ -\chi_{\kappa\kappa'}\biggl( -\nu\tilde{\eta}_{-\kappa}(-q)+\sqrt{\frac{T}{\Omega N_z}}i\tilde{\phi}_{-\kappa}(-q) \biggr) \biggl( \nu \tilde{\eta}_{\kappa'}(q)+\sqrt{\frac{T}{\Omega N_z}}i\tilde{\phi}_{\kappa'}(q) \biggr) \nonumber \\
    &\qquad\qquad+ 4N_z\rho_{\kappa\kappa'}|\mathbf{q}|^2\tilde{\eta}_{-\kappa}(-q)\tilde{\eta}_{\kappa'}(q) \biggr] + \frac{T}{2\Omega N_z}\sum_{\kappa q}\frac{\tilde{\phi}_{-\kappa}(-q)\tilde{\phi}_\kappa(q)}{\tilde{V}_\kappa(\mathbf{q})}, \label{S-eq:Sf}
\end{align}
where,
\begin{align}
    \chi_{\kappa\kappa'} &= \frac{-T}{\Omega N_z}\sum_{\kappa_1\kappa_2 k}\operatorname{Tr}[G_{\kappa+\kappa_1,\kappa'+\kappa_2}(k)\sigma_3G_{\kappa_2\kappa_1}(k)\sigma_3], \label{S-eq:chi} \\
    \rho_{\kappa\kappa'} &= \frac{\hbar^2T}{4m\Omega N_z^2}\biggl( \sum_{\kappa''k} \operatorname{Tr}[G_{\kappa+\kappa'',\kappa'+\kappa''}(k)\sigma_3] + \sum_{\kappa_1\kappa_2k}\frac{\hbar^2|\mathbf{k}|^2}{2m}\operatorname{Tr}[G_{\kappa+\kappa_1,\kappa'+\kappa_2}(k)G_{\kappa_2\kappa_1}(k)] \biggr). \label{S-eq:D}
\end{align}

Since the two layers are identical indicating $|\Delta_0|=|\Delta_1|\equiv \Delta_d$, one can check that $\chi_{\kappa\kappa'}=\chi_\kappa\delta_{\kappa\kappa'}$ and $\rho_{\kappa\kappa'}=\rho_\kappa\delta_{\kappa\kappa'}$ are all diagonal.
$\chi_\kappa$ is the static density susceptibility (number compressibility), and $\rho_\kappa$ is the stiffness of the physical pair-phase combination $2\sqrt{N_z}\eta_\kappa$. The factor $4N_z$ in Eq.~\eqref{S-eq:Sf} converts this stiffness to the normalized fermion-phase field; it is not an additional physical enhancement. The production calculation uses the gradient--gradient coefficient $\rho_\pi^{gg}$ and the density--density response $\chi_\pi$. This local-response approximation is valid in the weak interlayer-tunneling regime. Retaining all tunneling powers in $F_0$ does not remove this separate approximation. Their closed expressions and cutoff contributions follow below.

\subsection{Closed response kernels and internal cutoff contours}
\label{S-sec:fluct_coeffs}
Here we carry out the computation of the diagonal elements of Eqs.~(\ref{S-eq:chi}, \ref{S-eq:D}).
In this section only, we introduce the following shorthand notations to streamline the presentation of the formulae: \(z \equiv i\omega\), \(\xi \equiv \xi_{\mathbf{k}}\), \(\xi_{\pm} \equiv \xi \pm t\), \(E_{\pm} \equiv E_{0(\pi)}(\mathbf{k})\), $E_\pm' \equiv \partial E_\pm^2/\partial\xi$, \(\Delta_{\pm} \equiv \tilde{\Delta}_{0(\pi)}(\mathbf{k})\), and \(\Delta_{l} \equiv \Delta_{l}(\mathbf{k})\), such that \(\Delta_{\pm} = (\Delta_{0} \pm \Delta_{1})/2\).
First invert Eq.~(\ref{S-eq:G}) to obtain the blocks of the mean-field Green's function,
\begin{align}
    G_{00}(k) ={}& \frac{1}{(z^2-E_+^2)(z^2-E_-^2)} \nonumber \\
    &\times \begin{pmatrix} (z^2-\xi_+^2-|\Delta_+|^2)(z+\xi_-)-(z+\xi_+)|\Delta_-|^2 & (z^2-\xi_+^2-|\Delta_+|^2)\Delta_++\Delta_+^*\Delta_-^2 \\ (z^2-\xi_+^2-|\Delta_+|^2)\Delta_+^*+\Delta_+\Delta_-^{*2} & (z^2-\xi_+^2-|\Delta_+|^2)(z-\xi_-)-(z-\xi_+)|\Delta_-|^2 \end{pmatrix}, \\
    G_{\pi\pi}(k) ={}& [\text{$G_{00}(k)$ with $\xi_+$ and $\xi_-$ exchanged}], \\
    G_{0\pi}(k) ={}& \frac{1}{(z^2-E_+^2)(z^2-E_-^2)}\begin{pmatrix} (z+\xi_+)\Delta_+\Delta_-^*+(z+\xi_-)\Delta_+^*\Delta_- & [(z-\xi_+)(z+\xi_-)-|\Delta_-|^2]\Delta_-+\Delta_+^2\Delta_-^* \\ [(z+\xi_+)(z-\xi_-)-|\Delta_-|^2]\Delta_-^*+\Delta_+^{*2}\Delta_- & (z-\xi_+)\Delta_+^*\Delta_-+(z-\xi_-)\Delta_+\Delta_-^* \end{pmatrix}, \\
    G_{\pi 0}(k) ={}& [\text{$G_{0\pi}^\dagger(k)$ treating $z$ as real}].
\end{align}
The next step is plugging the Green's function into Eqs.~(\ref{S-eq:chi}, \ref{S-eq:D}) and do the algebra, including the frequency summation over $z$.
The diamagnetic contribution is combined with the current response using integration by parts. For a smooth kernel $X_{\mathbf k}$, the relevant identity is
\begin{equation}
\int_{\mathcal D}d^2k\,X_{\mathbf k}\nabla_{\mathbf k}^2\xi_{\mathbf k}
=\oint_{\partial\mathcal D}X_{\mathbf k}\nabla_{\mathbf k}\xi_{\mathbf k}\cdot d\mathbf S
-\int_{\mathcal D}d^2k\,\nabla_{\mathbf k}X_{\mathbf k}\cdot\nabla_{\mathbf k}\xi_{\mathbf k}.
\end{equation}
We use the bulk response with the internal pairing-window contour contributions specified below. An outer continuum cutoff is not a periodic lattice Brillouin-zone boundary; boundary contributions must therefore be handled consistently rather than discarded on grounds of lattice periodicity.
The algebra is cumbersome but straightforward, and the final results are
\begin{align}
    \chi_{0(\pi)}&=\chi_a\pm\chi_b, \label{S-eq:chi1} \\
    \chi_a &= \frac{1}{2\Omega}\sum_\mathbf{k} \biggl\{ \biggl[ \frac{2E_+^2-2\xi^2-|\Delta_0|^2-|\Delta_1|^2}{E_+^2-E_-^2} + \frac{4\xi^2(2E_+^2-2\xi^2+2t^2-|\Delta_0|^2-|\Delta_1|^2)}{(E_+^2-E_-^2)^2} - \frac{C_+(5E_+^2-E_-^2)}{E_+^2(E_+^2-E_-^2)^3} \biggr] \frac{f_+}{E_+} \nonumber \\
    &\qquad\qquad\quad+ \frac{C_+}{E_+^2(E_+^2-E_-^2)^2} f_+' + (E_+ \leftrightarrow E_-) \biggr\}, \label{S-eq:chi2} \\
    \chi_b &= \frac{1}{\Omega}\sum_\mathbf{k} t^2 \biggl\{ \biggl[ \frac{E_+^2 - E_-^2 + 8\xi^2}{(E_+^2-E_-^2)^2}-\frac{\xi^2(4E_+^2-|\Delta_0+\Delta_1|^2)(5E_+^2-E_-^2)}{E_+^2(E_+^2-E_-^2)^3} \biggr] \frac{f_+}{E_+} \nonumber \\
    &\qquad\qquad\quad+ \frac{\xi^2(4E_+^2-|\Delta_0+\Delta_1|^2)}{E_+^2(E_+^2-E_-^2)^2} f_+' + (E_+ \leftrightarrow E_-) \biggr\}, \label{S-eq:chi3} \\
    \rho_{0(\pi)}&=\rho_a\pm \rho_b, \label{S-eq:D1} \\
    \rho_a &= \frac{\hbar^2}{8N_zm\Omega}\sum_\mathbf{k} \frac{\hbar^2|\mathbf{k}|^2}{2m} \biggl\{ \biggl[ \frac{2E_+^2-6\xi^2+2t^2-|\Delta_0|^2-|\Delta_1|^2+2\xi E_+'}{E_+^2-E_-^2} - \frac{\xi(2E_+^2-2\xi^2+2t^2-|\Delta_0|^2-|\Delta_1|^2)}{E_+^2-E_-^2} \nonumber \\
    &\qquad\qquad\qquad\qquad\qquad\times \biggl( \frac{E_+'}{2E_+^2} + \frac{E_+' - E_-'}{E_+^2-E_-^2} \biggr) - \frac{C_+'}{(E_+^2-E_-^2)^2} + \frac{C_+''(5E_+^2-E_-^2)}{2E_+^2(E_+^2-E_-^2)^3}  \biggr] \frac{f_+}{E_+} \nonumber \\
    &\qquad\qquad\qquad\qquad\quad + \biggl[\frac{\xi(E_+^2-\xi^2-|\Delta_1|^2+t^2)E_+'}{E_+^2(E_+^2-E_-^2)} - \frac{C_+''}{2E_+^2(E_+^2-E_-^2)^2} \biggr] f_+' + (E_+ \leftrightarrow E_-) \biggr\}, \label{S-eq:D2} \\
    \rho_b &= \frac{\hbar^2}{4N_zm\Omega} \sum_\mathbf{k} \frac{\hbar^2|\mathbf{k}|^2t^2}{2m} \biggl\{ \biggl[ \frac{-1}{E_+^2-E_-^2} + \frac{(4\xi^2+|\Delta_0-\Delta_1|^2)(3E_+^2+E_-^2)}{(E_+^2-E_-^2)^3} \biggr] \frac{f_+}{E_+} \nonumber \\
    &\qquad\qquad\qquad\qquad\quad- \frac{4\xi^2+|\Delta_0-\Delta_1|^2}{(E_+^2-E_-^2)^2} f_+' + (E_+ \leftrightarrow E_-) \biggr\}. \label{S-eq:D3}
\end{align}
where
\begin{align}
    f_+ ={}& \tanh\biggl( \frac{E_+}{2T} \biggr),~f_+'\equiv \frac{\partial f_+}{\partial E_+} = \frac{1}{2T}\operatorname{sech}^2\biggl(\frac{E_+}{2T}\biggr), \\
    C_+ ={}& \xi^2[2(E_+^2-\xi^2+t^2)^2-2(E_+^2-\xi^2+t^2)(|\Delta_0|^2+|\Delta_1|^2)+|\Delta_0|^4+|\Delta_1|^4] , \\
    C_+' ={}& 2(E_+^2-\xi^2-t^2)(3E_+^2+\xi^2-t^2)-2(E_+^2+\xi^2-t^2)(|\Delta_0|^2+|\Delta_1|^2) \nonumber \\
    &+ (|\Delta_0|^2-|\Delta_1|^2)^2-2|\Delta_0\Delta_1|^2+4t^2(2\xi^2-\operatorname{Re}[\Delta_0\Delta_1^*]), \\
    C_+'' ={}& 2(E_+^2+\xi^2)(E_+^2-\xi^2-t^2)^2+4t^2(E_+^2-\xi^2)(2\xi^2-\operatorname{Re}[\Delta_0\Delta_1^*]) - 4(E_+^2-\xi^2)|\Delta_0\Delta_1|^2 \nonumber \\
    &- [2(E_+^2+\xi^2)(E_+^2-\xi^2-t^2) + 2t^2(2\xi^2-\operatorname{Re}[\Delta_0\Delta_1^*]) - (E_+^2-\xi^2)^2-t^4](|\Delta_0|^2+|\Delta_1|^2) \nonumber \\
    &+ (E_+^2+\xi^2)(|\Delta_0|^4+|\Delta_1|^4) + |\Delta_0|^2|\Delta_1|^4 + |\Delta_0|^4|\Delta_1|^2.
\end{align}

$(E_+ \leftrightarrow E_-)$ means an expression obtained by exchanging $E_+$ and $E_-$ in the expression ahead.
This complexity of the fluctuation coefficients stems from the inclusion of all orders of Josephson coupling energies.

The sharp pairing cutoff $w_\mathbf{k}=\Theta(\varepsilon_c-|\xi_\mathbf{k}|)$ makes $\Delta_l(\mathbf{k})$ discontinuous at the two internal contours $\xi_\mathbf{k}=s\varepsilon_c$, with $s=\pm1$.  Consequently, the integration by parts leading to Eq.~(\ref{S-eq:D2}) contains a small internal-boundary contribution in addition to the bulk term.  The correction is common to $\rho_0$ and $\rho_\pi$ and can be written as
\begin{equation}
    \delta \rho_{\mathrm{cut}} = \sum_{s=\pm1}\frac{s(\mu+s\varepsilon_c)}{8\pi N_z^2}
    \left\langle X_{\mathrm{SC}}(s\varepsilon_c,\alpha)-X_{\mathrm{N}}(s\varepsilon_c)\right\rangle_\alpha,
    \label{S-eq:Dcut}
\end{equation}
where $\langle A\rangle_\alpha=(2/\pi)\int_0^{\pi/2}d\alpha\,A(\alpha)$ and
\begin{align}
    X_{\mathrm{SC}}(\xi,\alpha) &=-\frac{1}{2} \left[
    \frac{E_+'}{E_+}\tanh\left(\frac{E_+}{2T}\right) + (E_+ \leftrightarrow E_-) \right], \\
    X_{\mathrm{N}}(\xi) &=-\tanh\left(\frac{\xi+t}{2T}\right)
    -\tanh\left(\frac{\xi-t}{2T}\right).
\end{align}
In $X_{\mathrm{SC}}$, the gap is evaluated on the paired side of the cutoff contour.  
We add Eq.~(\ref{S-eq:Dcut}) to both stiffnesses after evaluating the bulk kernels in Eqs.~(\ref{S-eq:D2}) and (\ref{S-eq:D3}).

When $\Delta_d=0$ (normal state), Eq.~(\ref{S-eq:Dcut}) vanishes because $X_{\mathrm{SC}}=X_{\mathrm{N}}$.

\subsection{Normal-state cancellation of the relative stiffness}
\label{S-sec:normal-state}
For zero gaps, the momentum-resolved overall stiffness kernel vanishes. The relative gradient--gradient kernel need not vanish pointwise, because tunneling couples the two layers. Its normal-state expression is
\begin{equation}
\rho_\pi\big|_{\Delta_0=\Delta_1=0}
=\frac{\hbar^2}{16N_zm\Omega}\sum_{\mathbf k}\frac{\hbar^2k^2}{2m}
\left[-\frac1t\tanh\!\left(\frac{\xi_{\mathbf k}+t}{2T}\right)+\frac{1}{2T}\operatorname{sech}^2\!\left(\frac{\xi_{\mathbf k}+t}{2T}\right)+(t\to-t)\right].
\label{S-eq:Dpi}
\end{equation}
The notation $(t\to-t)$ changes the sign of every occurrence of $t$ in the preceding two terms. Define $h(x)=\operatorname{sech}^2(x/2T)/(2T)$. The square bracket becomes
\begin{equation}
\mathcal B(\xi)=h(\xi+t)+h(\xi-t)-\frac1t\int_{-t}^{t}ds\,h(\xi+s).
\end{equation}
The two-dimensional parabolic density of states is constant, and $\hbar^2k^2/(2m)=\xi+\mu$. The integral is therefore proportional to $\int d\xi\,(\xi+\mu)\mathcal B(\xi)$. If $H_n=\int_{-\infty}^{\infty}d\xi\,\xi^nh(\xi)$, translating the integration variable gives
\begin{align}
\int_{-\infty}^{\infty}d\xi\,\mathcal B(\xi)
&=2H_0-\frac1t\int_{-t}^{t}ds\,H_0=0,\\
\int_{-\infty}^{\infty}d\xi\,\xi\mathcal B(\xi)
&=2H_1-\frac1t\int_{-t}^{t}ds\,(H_1-sH_0)=0.
\end{align}
Both moments cancel, so the normal-state polar integral of $\rho_\pi^{gg}$ vanishes even though its summand is nonzero. This proof uses the usual continuum extension of the energy integral to the real line. Corrections from physical band edges are exponentially small when those edges are far from $\xi=\pm t$ on the temperature scale. An artificial square momentum boundary can instead leave a residual. The internal pairing-cutoff correction vanishes separately at zero gap because $X_{\mathrm{SC}}=X_{\mathrm N}$; it is not responsible for the cancellation. Near zero amplitude, the calculated relative stiffness behaves as $\rho_\pi\propto\Delta_d^2$.

The stiffnesses reported for the fluctuating state are these microscopic kernels evaluated at the order parameter selected by SCHA. Within the retained quadratic gradient sector there is no extra spatial anharmonic renormalization. This definition differs from a helicity modulus including vortex, amplitude, or nonlocal response corrections. Dynamical anharmonic phase vertices, omitted here, are also required for a consistent beyond-Gaussian stiffness renormalization, including restoration of the continuum Galilean-invariance constraint~\cite{benfatto04lowenergy}. In the decoupled-layer limit, $\rho_\pi=\rho_0=\rho_{\mathrm{1L}}/2$, where $\rho_{\mathrm{1L}}$ is one isolated layer's pair-phase stiffness; the factor $1/2$ follows from using the physical phase sum and difference rather than individual layer phases.

\section{Self-consistent variational treatment and numerical methods}
\label{S-sec:methods}
\subsection{Gaussian trial theory and physical phase normalization}
\label{S-sec:gaussian-trial}
SCHA is most appropriate in low-temperature, phase-locked regimes with a well-developed pairing amplitude, where smooth long-wavelength phase modes dominate and vortices and chiral domain walls are dilute~\cite{benfatto01phase,liu23charge4e}. 
It incorporates Josephson anharmonicity self-consistently through an optimized Gaussian trial action, rather than assuming noninteracting bare fluctuations. 
SCHA-based calculations can yield good, model-dependent estimates of transition temperatures~\cite{you12tuning,moura23theoretical,yang26selfconsistent,newrock00two,rojas96critical}.

The local phase-dependent contribution to the action is
\begin{equation}
S_f''=\int_0^\beta d\tau\int d^2r\,
\left[V_J(\varphi+2\sqrt{N_z}\eta_\pi)-V_J(\varphi)\right].
\end{equation}
A one-loop treatment replaces this contribution by a quadratic potential with the bare Josephson mass $r_J^{\mathrm{bare}}=\partial_\varphi^2F_0$ for the physical relative phase. At configurations with negative bare mass, the low-momentum Gaussian mode is unstable and its determinant ceases to define a real equilibrium free energy. Setting a genuinely negative bare mass to zero does not resolve that instability. Polynomial truncations through fourth order are also unreliable when large phase fluctuations explore the periodic potential.

Instead, we represent $V_J$ by Fourier harmonics computed from the full mean-field spectrum and introduce a Gaussian trial potential $r_J(\delta\varphi)^2/2=2N_zr_J\eta_\pi^2$, with positive trial Josephson mass $r_J$. The symbol $r_J$ in the equations below refers to that trial mass unless explicitly marked as bare. Here mass denotes the quadratic phase-action coefficient, with units of energy per area, not the collective-mode gap. Before mass selection, the variational functional is $F_{\mathrm{SCHA}}(\Delta_d,\varphi,r_J;T,t,\theta)$.

Throughout the manuscript, $\rho_\kappa$ and $r_J$ refer to the physical pair-phase normalization. The normalized fermion-phase field used in the microscopic derivation is related to the physical relative phase and the notation of the main-text phase model by
\begin{equation}
\delta\varphi=2\sqrt{N_z}\eta_\pi,\qquad
W=\langle(\delta\varphi)^2\rangle_G=4N_z\langle\eta_\pi^2\rangle_G,\qquad
\rho_-=\rho_\pi,\qquad
\chi_-(\mathbf q)=\frac{1}{4N_z[\chi_\pi^{-1}+\widetilde V_\pi(\mathbf q)]}.
\label{S-eq:physicalphasemap}
\end{equation}
The collective-mode gap obeys $\Omega_J^2=r_J/\chi_-(\mathbf 0)$, as defined in End Matter. The bare density response $\chi_\pi$ and screened temporal susceptibility $\chi_-(\mathbf q)$ are distinct quantities; $\kappa$ is reserved for the layer Fourier index.

The archived production arrays retain the field names \texttt{D\_0}, \texttt{D\_pi}, and \texttt{D\_J\_eff}; their values are $4N_z\rho_0$, $4N_z\rho_\pi$, and $4N_zr_J$, respectively. Thus the physical stiffnesses and trial mass are the stored values divided by $4N_z=8$. The numerical floor in Eq.~\eqref{S-eq:floormain} is the unchanged stored floor divided by the same factor. The archived \texttt{eta\_pi2} field stores $\langle\eta_\pi^2\rangle_G=W/(4N_z)$, not the physical phase variance $W$ itself. No spectrum or free energy is changed by this notation conversion.

With this trial potential, integrating out phase and density fields gives the spectra
\begin{equation}
\omega_0(\mathbf q)=\sqrt{4N_z\rho_0|\mathbf q|^2[\chi_0^{-1}+\widetilde V_0(\mathbf q)]},\qquad 
\omega_\pi(\mathbf q)=\sqrt{4N_z(\rho_\pi |\mathbf q|^2+r_J)
[\chi_\pi^{-1}+\widetilde V_\pi(\mathbf q)]}.
\label{S-eq:modesmain}
\end{equation}
The associated Gaussian free energy and equal-time physical relative-phase variance $W=\langle(\delta\varphi)^2\rangle_G=4N_z\langle\eta_\pi^2\rangle_G$, as in the main-text phase model, are
\begin{align}
F_G(r_J;T,t)&=\intq\sum_{\kappa=0,\pi}
\left[T\ln(1-e^{-\omega_\kappa/T})+\frac{\omega_\kappa}{2}\right],
\label{S-eq:FGmain}\\
W(r_J;T,t)&=\intq\frac{1}{2\chi_-(\mathbf q)\omega_\pi(\mathbf q)}
\coth\!\left(\frac{\omega_\pi(\mathbf q)}{2T}\right).
\label{S-eq:Wmain}
\end{align}
The momentum cutoff is $\Lambda=\Delta_d/(\hbar v_F)$, with $v_F=\sqrt{2\mu/m}$, excluding fluctuations shorter than the coherence-length scale. The factor $1/2$ in the zero-point energy counts the real phase field correctly.
The quoted angular windows and temperature ratios describe this effective model, not precise material predictions. Quadrature and mass-floor convergence do not establish independence from the physical cutoff or control of omitted nonlocal response.

\subsection{Variational free energy and self-consistent mass}
\label{S-sec:variational-mass}
At fixed amplitude, stationary relative phase, and microscopic gradient and density coefficients, the Feynman--Bogoliubov inequality for the specified phase action reads~\cite{feynman18statistical,fishman88role,benfatto01phase,yang26superconducting}
\begin{equation}
F\leq F_{\mathrm{SCHA}}=F_0+F_G
+\langle V_J(\varphi+2\sqrt{N_z}\eta_\pi)\rangle_G
-V_J(\varphi)-\frac{r_JW}{2}.
\label{S-eq:F}
\end{equation}
The subtraction is essential. $F_0$ already contains $V_J(\varphi)$, while $F_G$ already contains the trial harmonic energy. Retaining either twice would change the variational stationarity condition.

For the real Gaussian field $\eta_\pi$, $\delta\varphi=2\sqrt{N_z}\eta_\pi$ has variance $W$. Gaussian averaging therefore gives
\begin{equation}
\left\langle e^{in\delta\varphi}\right\rangle_G=e^{-n^2W/2}.
\end{equation}

Equivalently, in terms of the Josephson Fourier coefficients,
\begin{equation}
F_{\mathrm{SCHA}}(\Delta_d,\varphi,r_J;T,t,\theta)=F_0+F_G-\frac{r_JW}{2}
+\sum_{n=1}^{N_J}J_n\cos(n\varphi)
\left(e^{-n^2W/2}-1\right).
\label{S-eq:Fvarmain}
\end{equation}

At fixed microscopic response coefficients,
\begin{equation}
\frac{\partial F_G}{\partial r_J}=\frac W2,
\qquad
\frac{\partial\langle V_J\rangle_G}{\partial W}
=\frac12\langle V_J''\rangle_G.
\end{equation}
Differentiating Eq.~\eqref{S-eq:F} yields
\begin{equation}
\frac{\partial F_{\mathrm{SCHA}}}{\partial r_J}
=\frac12\left[\langle V_J''\rangle_G-r_J\right]
\frac{\partial W}{\partial r_J}.
\end{equation}
Thus the stationary mass obeys
\begin{equation}
r_J=-\sum_{n=1}^{N_J}n^2J_n\cos(n\varphi)
e^{-n^2W/2},
\label{S-eq:SCHAmass}
\end{equation}
together with the variance in Eq.~\eqref{S-eq:Wmain}. The factors of $N_z$ enter through $W=4N_z\langle\eta_\pi^2\rangle_G$, not through a rescaling of the physical mass or Josephson coefficients.

The trial mass generally differs from the curvature of the final amplitude-relaxed free energy. Differentiating that free energy also differentiates the variance, microscopic response coefficients, and relaxed order parameter. A mass approaching zero at a single phase therefore does not establish that the phase landscape is flat in a neighborhood.

\subsection{Periodic averaging of a compact phase}
\label{S-sec:compact-phase}
A Gaussian field on the real line can be used to average a periodic potential without restricting its unwrapped variance. Let $x$ denote the unwrapped relative-phase fluctuation with variance $W$, and define the wrapped probability density of $\vartheta\in[-\pi,\pi)$ by
\begin{equation}
P(\vartheta)=\frac{1}{\sqrt{2\pi W}}
\sum_{m\in\mathbb Z}\exp\!\left[-\frac{(\vartheta+2\pi m)^2}{2W}\right].
\label{S-sec:scha_periodic}
\end{equation}
It is periodic and normalized over one period. Poisson summation gives
\begin{equation}
P(\vartheta)=\frac{1}{2\pi}\sum_{n\in\mathbb Z}e^{-n^2W/2+in\vartheta}
=\frac{1}{2\pi}\left[1+2\sum_{n=1}^\infty e^{-n^2W/2}\cos(n\vartheta)\right].
\end{equation}
For any $2\pi$-periodic function $A$,
\begin{equation}
\int_{-\pi}^{\pi}d\vartheta\,P(\vartheta)A(\vartheta)
=\int_{-\infty}^{\infty}\frac{dx}{\sqrt{2\pi W}}e^{-x^2/(2W)}A(x).
\end{equation}
The Fourier factors $e^{-n^2W/2}$ are exactly the Debye--Waller factors in the mass equation. As the unwrapped variance grows, $P$ approaches $1/(2\pi)$; there is no inconsistency with compactness. This construction justifies Gaussian averages of the local periodic potential. It does not incorporate spatial vortices or domain-wall fluctuations and hence does not remove the critical limitations of SCHA.

\subsection{Harmonic competition and restoration of time-reversal symmetry}
\label{S-sec:harmonic-competition}
A local measure of chirality is $\mathcal C\propto\langle\sin(\varphi+\delta\varphi)\rangle$. In a Gaussian state centered on $\varphi$,
\begin{equation}
\mathcal C\propto e^{-W/2}\sin\varphi.
\end{equation}
An equally weighted distribution around opposite chiral phases has $\mathcal C=0$, but need not be localized around zero phase. Symmetry restoration alone therefore does not explain a preference for a locked achiral configuration.

To isolate the energetic mechanism, consider $V_J=-A\cos\varphi+B\cos2\varphi$, with $A,B>0$, at fixed amplitude and variance. Gaussian averaging gives $A_{\mathrm{eff}}=Ae^{-W/2}$ and $B_{\mathrm{eff}}=Be^{-2W}$. Its phase derivative is
\begin{equation}
\partial_\varphi\langle V_J\rangle_G
=\sin\varphi\left(A_{\mathrm{eff}}-4B_{\mathrm{eff}}\cos\varphi\right).
\end{equation}
The chiral stationary phases consequently satisfy
\begin{equation}
\cos\varphi_0=\frac{A}{4B}e^{3W/2}<1,
\end{equation}
and merge at zero phase when this ratio reaches unity. For a larger ratio, zero phase has positive local curvature, $A_{\mathrm{eff}}-4B_{\mathrm{eff}}>0$. Thus differential harmonic suppression can favor a coherent achiral state away from $45^\circ$. The full calculation retains eight harmonics and determines the amplitude, variance, and microscopic response coefficients self-consistently, so this fixed-variance example is illustrative rather than a substitute for minimizing Eq.~\eqref{S-eq:F}.

The analytic energy scale in the main text follows by completing the square at fixed amplitude and variance. Defining $r_{\mathrm{eff}}=A_{\mathrm{eff}}/(4B_{\mathrm{eff}})$ gives
\begin{equation}
\langle V_J^{(2)}\rangle_G
=2B_{\mathrm{eff}}(\cos\varphi-r_{\mathrm{eff}})^2
-B_{\mathrm{eff}}-\frac{A_{\mathrm{eff}}^2}{8B_{\mathrm{eff}}}.
\end{equation}
For $0\leq r_{\mathrm{eff}}<1$, its chiral minima have $\cos\varphi_0=r_{\mathrm{eff}}$, and their energy density relative to the neighboring achiral configuration is $\delta f_\chi=2B_{\mathrm{eff}}(1-r_{\mathrm{eff}})^2$. The corresponding trial mass is
\begin{equation}
r_J=\partial_\varphi^2\langle V_J^{(2)}\rangle_G\big|_{\varphi_0}
=4B_{\mathrm{eff}}(1-r_{\mathrm{eff}}^2).
\end{equation}
Both vanish when the minima merge at $r_{\mathrm{eff}}=1$. At $45^\circ$, $A=0$, giving Eq.~\eqref{eq:chiralscale} in End Matter after the normalization conversion in Eq.~\eqref{S-eq:physicalphasemap}. These are local Josephson energy and mass scales at fixed variance, not derivatives of the fully relaxed variational free energy. Multiplication by a specified coherent area gives an energy, but identifying it directly with $k_BT_\chi$ neglects spatial correlations and possible defects.

At exactly $45^\circ$, symmetry requires $V_J(\varphi+\pi)=V_J(\varphi)$ and hence $J_{2m+1}=0$. For a dominant positive $J_2=B$, the averaged potential remains $Be^{-2W}\cos2\varphi$. Its minima stay at $\pm\pi/2$ for any finite variance. The mass equation at these minima reduces to $r_J=4Be^{-2W}$; at zero phase its right-hand side instead has the opposite sign. Suppressing this harmonic cannot by itself generate a positive-mass zero-phase state. Higher even harmonics are included in the numerical calculation, but do not lift the symmetry prohibition on odd harmonics.

For finite temperature and nonzero relative spatial stiffness, the continuum variance diverges logarithmically as $r_J\to0^+$. All local Josephson harmonics then vanish in this limiting Gaussian average, and the center phase no longer diagnoses coherent locking. This thermal infrared divergence is absent at zero temperature in two dimensions. The small positive mass regulator used numerically need not give a large variance in practice; its finite value does not establish a uniform phase distribution or a thermodynamic phase boundary. Establishing spatial relative-phase disorder requires correlation functions or a treatment of defects, not merely the center phase evaluated with this regulator.

\subsection{Branch selection and limits of the approximation}
\label{S-sec:branch-selection}
The scalar mass equation can have more than one positive root. The numerical prescription searches from high mass and selects the largest resolved positive root continuously connected to the locked state. The outer minimization is performed on this branch. The Gaussian trial family also includes the zero-mass boundary $r_J\to0^+$; its admissibility does not imply that it is stationary or minimizing. When no positive root is bracketed above the numerical floor, this boundary is evaluated with the small positive regulator
\begin{equation}
r_J^{\mathrm{floor}}=1.25\times10^{-13}\,
\frac{\mathrm{meV}}{(10\,\mathrm{nm})^2},
\label{S-eq:floormain}
\end{equation}
and flagged as floor limited. This regulator is solely a numerical guard against singular evaluations, not a physical Josephson gap or an additional self-consistent locked solution. Its value is the unchanged production floor divided by $4N_z=8$ to use the physical pair-phase normalization in Eq.~\eqref{S-eq:physicalphasemap}. Adaptive fluctuation-momentum quadrature uses absolute and relative tolerances $10^{-5}$ for the coarse search and $1.49\times10^{-8}$ for final evaluation. Features that depend on the finite regulator are not by themselves physical predictions.

This prescription avoids arbitrary jumps among trial-mass roots, but it does not compare all stationary variational branches. Loss of the locked root is a spinodal criterion, while an equilibrium phase boundary requires comparison of competing free-energy minima. For the current-biased calculation, we use the same mass floor for the boundary branch and include it when one-sided stable; loss of the first static branch need not imply loss of every reachable static state (Sec.~\ref{S-sec:current-bias}). The variational upper bound on $F$ alone does not prove an upper bound on the physical transition temperature. 

\subsection{Momentum quadrature and harmonic projection}
\label{S-sec:quadrature-projection}
Electronic integrals use polar coordinates $\bar k=ka$ over the disk $0\leq\bar k<\sqrt{2}\pi$. Fourfold symmetry reduces the angular interval to $0\leq\alpha<\pi/2$. Composite Gauss--Legendre radial integration is split at every accessible feature $\xi=-\varepsilon_c,-|t|,0,|t|,\varepsilon_c$, and at the two domain endpoints. At the adopted resolutions, each interval receives at least eight radial points, with the remaining points distributed by interval width. The angle is sampled by a periodic midpoint rule. Meshes are rebuilt when tunneling changes, and momentum sums are distributed using MPI.

Josephson harmonics are computed from
\begin{equation}
J_n(\Delta_d;T,t,\theta)=\frac{1}{\pi}\int_0^{2\pi}
F_0(\Delta_d,\varphi;T,t,\theta)\cos(n\varphi)\,d\varphi,
\qquad n\geq1.
\label{S-eq:projection}
\end{equation}
We sample 32 phases during the coarse search and 64 during final polishing and verification, retaining eight harmonics. A real FFT with forward normalization gives half the cosine coefficient at positive frequencies. Local Fourier coefficients are combined by a single MPI array reduction. This is a truncated Fourier representation of the full microscopic phase dependence, rather than a fourth-order phase or tunneling expansion.

\subsection{Global minimization and phase-dependent observables}
\label{S-sec:global-minimization}
Each parameter point is searched independently using SciPy~\cite{2020SciPy-NMeth}. Differential evolution and initial local polishing use 300 radial by 256 angular points. Bounded Nelder--Mead refinement uses 600 by 512 points. The three lowest distinct candidates are reevaluated on a 1200 by 1024 mesh before selecting the lowest energy. Seeds include the mean-field solution, differential-evolution candidates, a real-component state, and an equal-component state. The tunneling scan additionally samples 17 amplitudes on the achiral line and polishes the three lowest samples. Mean-field references use BFGS with an analytical gradient on the 600 by 512 mesh.

Main-text Fig.~\ref{fig:landscape} uses independently amplitude-relaxed phase profiles at $T/\TcL=0.245\%$. At each fixed $\varphi$, a global amplitude search on $0\leq\Delta_d\leq80$ meV combines SciPy's simplicial homology global optimization (SHGO) with adaptive sampling and bounded Nelder--Mead polishing of candidate minima. The initial grid contains 33 uniformly spaced amplitudes, supplemented by logarithmically spaced offsets around every accessible zero-phase node-merger amplitude, $|t/\sin\theta|$ and $|t/\cos\theta|$, on their thermal scales. Neighboring intervals of sampled minima and regions of large energy curvature are bisected before global optimization. Every trial is evaluated on the prescribed mass branch, including candidates far from the previously selected amplitude. This replaces the earlier coarse amplitude search, which missed a narrow lower minimum near $\theta=22.5^\circ$. Independent searches starting with 65 uniform amplitudes and additional adaptive refinement test amplitude completeness numerically; they do not constitute a proof for arbitrary unsampled minima.

Composite Gauss--Legendre fermion quadrature resolves the nodes at $\varphi=0,\pi$ and their thermal neighborhoods. The interval boundaries follow the amplitude but not the imposed phase. Order six per interval is used throughout the landscape's global amplitude search, with orders eight and twelve used for independent checks. This resolution is important because narrow-feature slopes can remain sensitive to a uniform momentum mesh even after phase refinement.

The profiles start from 61 uniform phases over $[0,\pi]$, augmented by 15 logarithmic samples near each endpoint, down to $10^{-5}\pi$. Adaptive interval halving near narrow energy features, mass-branch boundaries, and amplitude switches repeats the global amplitude search at every added phase. The dashed floor-limited segments in Fig.~\ref{fig:landscape} visualize the regulated boundary, not an additional locked state. Figure~\ref{fig:current} instead uses the separate current-biased continuation in Sec.~\ref{S-sec:current-bias}; it is not extracted by maximizing derivatives of this regulated landscape.

Independent landscape checks refine the amplitude grid and fermion quadrature, and vary the physical mass floor from $1.25\times10^{-15}$ to $1.25\times10^{-11}$ meV$/(10\,\mathrm{nm})^2$, bracketing Eq.~\eqref{S-eq:floormain}. At representative angles, we also raise the amplitude bound to $120$ meV, double the Josephson projection to 128 phases, and disable reuse of phase-independent harmonic coefficients. Regulated floor energies can still vary; they are not differentiated through a branch boundary.

On a smooth internally stationary branch, the envelope theorem removes explicit derivatives of the optimized amplitude and mass from the first phase derivative. Changing the selected minimum nevertheless changes the evaluated phase slope. A crossing of amplitude minima can give a continuous lower envelope with finite, unequal one-sided derivatives. Such a variational cusp is distinct from loss of a positive mass root and does not by itself predict a device switching current. Numerical convergence does not establish the unrestricted Gaussian minimum or accuracy of omitted physics.

The minimum quasiparticle gap is obtained from the lower branch $E_-(\mathbf k)$. No statistical inference is involved; the curves and phase maps are deterministic numerical results for the stated model. The fluctuation calculations are restricted to $T/\TcL\leq50\%$, with the main conclusions drawn from the much lower-temperature chiral region.

\subsection{Adiabatic current-biased stability and recapture}
\label{S-sec:current-bias}
Main-text Fig.~\ref{fig:current}(a) reports the adiabatic static-current limit $j_c^{\mathrm{ad}}$ at $t=10$ meV and $T/\TcL=3\%,5\%,10\%$ on a linear scale. Our validation additionally includes $T/\TcL=0.5\%,2\%$. All curves use the same normalization $j_{\mathrm{ref}}=j_c^{\mathrm{MF}}(T=0,\theta=0)=4.12\times10^{11}$ A/m$^2$, evaluated with amplitude relaxation at the same $t$. This is a nonequilibrium current-biased stability problem, not the zero-current equilibrium phase diagram. The reference $\TcL=203.8$ K is the model's monolayer mean-field pairing scale, not an experimental transition temperature. We retain two trial-mass branches: the largest resolved positive root and the regulated boundary $r_J=r_J^{\mathrm{floor}}$, with the same floor as in Eq.~\eqref{S-eq:floormain}. Starting from the lowest-free-energy stable zero-current state, we increase the bias $j_b$ and follow a local minimum of
\begin{equation}
G_{j_b}(\varphi,\Delta_d,r_J;T,t,\theta)
=F_{\mathrm{SCHA}}(\Delta_d,\varphi,r_J;T,t,\theta)
-\frac{\hbar}{2e}j_b\varphi.
\label{S-eq:current-tilt}
\end{equation}
The stationary current obeys main-text Eq.~\eqref{eq:current}. A finite $r_J$ alone does not guarantee stability. Let $H$ be the Hessian of $F_{\mathrm{SCHA}}$ in $\varphi$ and the internal variables $x=(\Delta_d,r_J)$, evaluated at a stationary state. We require positive retained spatial and temporal response coefficients on both branches. On a positive-mass branch, we additionally require $H_{xx}>0$ and
\begin{equation}
K=H_{\varphi\varphi}-H_{\varphi x}H_{xx}^{-1}H_{x\varphi}>0.
\label{S-eq:current-stability}
\end{equation}
Here $K=d^2\overline F/d\varphi^2$ is the internally relaxed phase curvature; $K=0$ marks a phase fold when the internal block remains positive. The numerical implementation uses $\ln r_J$ for positive mass, which preserves these local stability signs at stationarity. The common-amplitude restriction is retained: this is a full Hessian test within the three uniform variational coordinates, not a test against arbitrary spatial or unequal-layer-amplitude perturbations.

At the regulated boundary, we evaluate the full functional in Eq.~\eqref{S-eq:Fvarmain} at fixed $r_J^{\mathrm{floor}}$, retaining the finite variance $W(r_J^{\mathrm{floor}};T,t)$ and its Debye--Waller factors. The floor is an active constraint, not a stationary positive-mass root. We reoptimize the common amplitude and phase and require
\begin{equation}
\left.\frac{\partial F_{\mathrm{SCHA}}}{\partial r_J}\right|_{r_J^{\mathrm{floor}}}
\geq0,\qquad H_{\Delta_d\Delta_d}>0,\qquad
K_{\mathrm{floor}}
=H_{\varphi\varphi}-\frac{H_{\varphi\Delta_d}^2}{H_{\Delta_d\Delta_d}}>0.
\label{S-eq:current-floor-stability}
\end{equation}
The mass inequality is strict at the reported boundary states, so the amplitude--phase Hessian determines stability along the constraint. Boundary currents differentiate the amplitude-relaxed free energy at this fixed regulator; no stencil crosses a mass-branch switch or a gap between separately resolved landscape segments.

For an independent limiting check, at finite $T$ in the infinite-area continuum, $W$ diverges as $r_J\to0^+$, while $r_JW\to0$ and every retained averaged Josephson harmonic vanishes. The strict zero-mass functional is
\begin{equation}
F_{\mathrm{SCHA}}^{[0]}(\Delta_d,\varphi;T,t,\theta)
=F_0-\sum_{n=1}^{N_J}J_n\cos(n\varphi)+F_G(0;T,t).
\label{S-eq:current-zero-mass}
\end{equation}
The superscript $[0]$ denotes the exact zero-mass boundary, not zero temperature. This expression is a limiting diagnostic, not the prescription for the plotted open diamonds. The local response coefficients entering $F_G$ retain unaveraged background-phase dependence even when the averaged Josephson harmonics vanish. A finite numerical floor regularizes the variance but does not make this approximation controlled for broad relative-phase fluctuations. Neither constrained stability nor weak floor dependence establishes a physical Josephson current in that regime; consistent phase averaging of the response and current vertices, finite-size effects, and nonlocal dynamics require further treatment. The open diamonds therefore remain provisional boundary estimates, not additional self-consistent locked solutions.

After the first fold or internal instability, we search both retained branches for a stable static state at the same bias. Recapture is accepted only when an internally stable downhill path in $G_{j_b}$ reaches that state. A positive-mass-to-floor change additionally requires a downhill mass-energy bridge ending at the finite floor, downhill amplitude relaxation, and a one-sided stable endpoint. We then continue the recaptured branch to its final static limit. This corresponds to adiabatically fast internal relaxation and positive phase mobility without activation; it does not simulate inertia, the circuit, noise, or a sustained voltage state. A transient phase jump can relax to a new static phase instead of producing a running phase. For example, at $\theta=13.5^\circ$ and $T/\TcL=0.5\%$, the first fold occurs at $3.91\times10^{10}$ A/m$^2$, but downhill recapture into a second positive-mass minimum permits currents up to $1.78\times10^{11}$ A/m$^2$. Identifying the first fold with the critical current would produce a spurious angular dip.

We check all 80 angle--temperature combinations on the 16-angle grid of Fig.~\ref{fig:current}(a), including the additional $0.5\%$ and $2\%$ benchmarks. Twenty require recapture, including four positive-mass-to-floor changes. Independent global amplitude searches retain all resolved minima as branch seeds, followed by adaptive amplitude sampling and targeted phase and endpoint refinement. A fold must be a resolved local current maximum with a correctly oriented change between stable and unstable states. Fermion quadrature orders six and eight are used for discovery and targeted global refinement, respectively, followed by order twelve for the final endpoints and continuation. The current calculation retains $N_J=32$--128 Josephson harmonics, with 128--512 projection phases, requiring endpoint-current agreement within $0.1\%$ between converged cutoffs. Independent order-sixteen quadrature and halved derivative steps change endpoint currents by less than $0.0001\%$. All reported terminal states and recapture paths pass the stated stability tests. These tests establish numerical consistency on the sampled grid, not completeness over all Gaussian branches or unresolved angular structure.

A larger static capacity in a disconnected amplitude basin is not automatically reachable. At $18^\circ$ and $T/\TcL=0.5\%$, an explicit full phase-slip cycle, including downhill transitions between positive mass and the floor, returns to the original amplitude minimum rather than the separate higher-current metastable minimum.

For every branch competition we use the same finite floor. At all eight floor-regulated endpoints we reoptimize the amplitude at physical floors $1.25\times10^{-15}$ and $1.25\times10^{-11}$ meV$/(10\,\mathrm{nm})^2$, a factor of 100 below and above the production floor. At $45^\circ$, the $T/\TcL=3\%,5\%,10\%$ endpoints are additionally reoptimized in phase and checked along their constrained stable branches. The largest current shift from the production value is $0.46\%$. The $45^\circ$ endpoints at $5\%$ and $10\%$ exceed the stricter $0.1\%$ regulator tolerance despite passing the numerical-resolution tests. At $3\%$, the $0^\circ$ and $45^\circ$ endpoints also exceed that stricter tolerance, with sensitivity of about $0.14\%$. This remaining regulator dependence is distinct from quadrature error; no floor-regulated estimate is promoted to a physical prediction by these checks.

The intermediate-angle ramp originates in the node-merger response described in Sec.~\ref{S-sec:current-node-merger}. It reaches $1.21\times10^{12}$ A/m$^2$ at the sampled angle $19.125^\circ$ for $T/\TcL=0.5\%$. At $3\%$ the intermediate-angle peak is $2.21\times10^{11}$ A/m$^2$ at $20.25^\circ$, while the feature is a small hump at $5\%$, and is absent on this grid at $10\%$. Neither its angle nor its disappearance temperature is universal. No phase-exclusion window is applied to remove the resolved narrow features.

There is also a distinct zero-twist endpoint enhancement: recapture onto the floor-regulated boundary gives $j_c^{\mathrm{ad}}=1.62\times10^{12}$ and $2.39\times10^{11}$ A/m$^2$ at $0.5\%$ and $3\%$, respectively. The spacing to the next sampled angle, $4.5^\circ$, does not resolve its width; it must not be interpreted as a resolved finite-width peak. At $45^\circ$, the terminal floor-regulated currents decrease monotonically with temperature, with plotted values $(2.49,2.32,1.97)\times10^{10}$ A/m$^2$ at $3\%,5\%,10\%$. Both observations are subject to the local-response caveat above. In particular, the finite-phase transport limit should not be inferred from the equilibrium plasma gap or identified with equilibrium chirality.

Some intermediate-angle points in Wang et al.~\cite{Wang2023NatCommun} and Zhao et al.~\cite{Zhao2023Science} show qualitatively similar enhancements. Main-text Fig.~\ref{fig:current}(b) redraws all 26 original 12 K points from the authors' \href{https://doi.org/10.7910/DVN/BYQDSP}{deposited dataset}, corresponding to Zhao et al.'s Fig.~2B. Assuming reflection symmetry, we plot $\theta=\min(\widetilde\theta,90^\circ-\widetilde\theta)$, where $\widetilde\theta=\theta_o\bmod90^\circ$ with $\theta_o$ the original twist angle. Filled squares and open circles distinguish $\widetilde\theta\leq45^\circ$ and $\widetilde\theta>45^\circ$ before folding; repeated folded angles remain separate points. Vertical bars reproduce the sample standard deviation of the available tabulated values at 9--15 K, following the deposited plotting procedure. The natural cubic smoothing spline uses a fixed penalty of $10\,\mathrm{deg}^3$, with equal weight per device and no extrapolation. It is only a guide to the eye, not a theoretical fit. These measurements concern finite-thickness flakes and report $I_cR_N$, the product of junction critical current $I_c$ and normal resistance $R_N$, rather than current density. Their temperature and junction parameters differ from those of our model. Establishing a common origin requires systematic temperature-dependent angular scans and junction-specific resistance and dynamical models; these observations do not confirm our mechanism or the $t^4$ chiral-ordering law.

\subsection{Node merger and enhanced phase response}
\label{S-sec:current-node-merger}
The intermediate-angle current enhancement is tied to a reconstruction of the nodal quasiparticle spectrum, not to equilibrium chiral order. Consider the zero-relative-phase background $\varphi=0$ at fixed common amplitude $\Delta_d>0$ and tunneling $t>0$. Within the pairing shell $w_{\mathbf k}=1$, write $\xi=\xi_{\mathbf k}$ and $d_{0,1}=\Delta_d\cos(2\alpha\pm\theta)$ for the real layer gaps, where $\alpha=\alpha_{\mathbf k}$. Equation~\eqref{S-eq:bands} implies
\begin{equation}
E_0^2E_\pi^2
=\bigl(\xi^2-t^2-d_0d_1\bigr)^2
+\xi^2(d_0+d_1)^2.
\label{S-eq:current-node-condition}
\end{equation}
Since $d_0d_1=\Delta_d^2[\cos^2(2\alpha)-\sin^2\theta]$, the nodes near $\alpha=\pi/4$, for $0<\theta\leq45^\circ$, belong to two families:
\begin{align}
\xi&=0,\quad
\cos^2(2\alpha_n)=\sin^2\theta-(t/\Delta_d)^2,
&&t\leq\Delta_d\sin\theta,
\nonumber\\
\cos(2\alpha_n)&=0,\quad
\xi^2=t^2-\Delta_d^2\sin^2\theta,
&&t\geq\Delta_d\sin\theta.
\label{S-eq:current-node-families}
\end{align}
The angularly separated nodes on $\xi=0$ merge at
\begin{equation}
t=\Delta_d\sin\theta_{\mathrm m},
\qquad \theta_{\mathrm m}=\arcsin(t/\Delta_d),
\label{S-eq:current-node-merger}
\end{equation}
and separate radially on the other side. Thus the zero-phase merger does not itself remove the nodes.

At the merger, define the energy coordinates $u=\xi$ and $v=2\Delta_d(\alpha-\pi/4)$, both linear in momentum displacement to leading order. Expanding Eq.~\eqref{S-eq:current-node-condition}, with $E_0\simeq2t$, gives
\begin{equation}
E_\pi^2\simeq
\frac{(u^2-v^2)^2+4u^2v^2\cos^2\theta_{\mathrm m}}{4t^2}.
\label{S-eq:current-quadratic-touching}
\end{equation}
The dispersion is therefore quadratic rather than Dirac-like. In the leading small-twist nodal approximation it reduces to $E_\pi\simeq(u^2+v^2)/(2t)$, the quadratic touching of the circular-Fermi-surface magic-angle theory~\cite{volkov23magic}. The low-energy density of states is finite at the touching, rather than vanishing linearly with energy as for an isolated Dirac cone. This enhanced spectral weight makes the response to a small phase-induced gap unusually strong.

At finite temperature, low-energy quasiparticles soften the superfluid stiffness and lower the overall-phase mode frequencies and zero-point energy. An imposed nonzero relative phase gaps these states~\cite{volkov23current}, suppressing their contribution to the softening. The recovery of stiffness raises the overall-phase zero-point energy rapidly away from zero phase, producing a narrow, continuous dip and a large phase derivative of the full variational free energy. In our calculation this collective contribution, together with amplitude relaxation, enhances the current; it is not simply an increase of a bare Josephson harmonic. Thermal rounding weakens the enhancement as temperature rises.

For a numerical scale check, the zero-current amplitude at the sampled $3\%$-temperature peak, $\theta=20.25^\circ$, is $\Delta_d\simeq29.08$ meV. With $t=10$ meV, Eq.~\eqref{S-eq:current-node-merger} gives $\theta_{\mathrm m}\simeq20.11^\circ$, close to that peak. This is a local consistency check, not an independent prediction of the peak position: the merger condition is defined at zero phase and fixed amplitude, whereas the critical current occurs at finite phase with amplitude relaxation. Both the merger angle and the current maximum depend on the model parameters.

\section{Weak-tunneling coefficients and numerical benchmarks}
\label{S-sec:weak-tunneling}
\subsection{Quantum variance in the decoupled limit}
\label{S-sec:decoupled-variance}
It is also important to distinguish a vanishing energy scale from the absence of a zero-temperature positive-mass solution. To illustrate the infrared behavior, take the screened temporal susceptibility $\chi_->0$ as momentum independent over the fluctuation window and keep $\rho_-=\rho_\pi>0$ fixed. The zero-temperature physical phase variance $W_{\mathrm q}$ is then
\begin{align}
W_{\mathrm q}(r_J)
&=\frac{1}{4\pi\rho_-\sqrt{\chi_-}}
\left[\sqrt{\rho_-\Lambda^2+r_J}-\sqrt{r_J}\right],
\label{S-eq:weakquantumvariance}\\
W_{\mathrm q}(0)&=\frac{\Lambda}{4\pi\sqrt{\rho_-\chi_-}}.
\end{align}
The finite limit also holds for the actual momentum-dependent Coulomb kernel, which is regular in the relative channel at $\mathbf q=\mathbf 0$. At $45^\circ$ and small $B$, the two-harmonic mass equation consequently has the fixed-background asymptote
\begin{equation}
r_J\sim4B\,e^{-2W_{\mathrm q}(0)}>0.
\label{S-eq:weakquantummass}
\end{equation}
The mass and Josephson stabilization energy tend to zero with $B$, but zero-point fluctuations alone do not force this root to disappear at a nonzero coupling in this limit. At finite temperature, the thermal infrared divergence changes the mass equation and an ordering transition requires a spatial analysis. At exactly $t=0$, all Josephson harmonics vanish and there is no interlayer phase locking. Neither the finite-$t$ scan nor a floor-limited mass establishes a universal law for $T_\chi$ as $t\to0^+$.

\subsection{Zero-temperature coefficient of the weak-tunneling splitting}
\label{S-sec:weak-splitting}
Here we make explicit the coefficient $C_Q$ in Eq.~\eqref{eq:CQmain} of End Matter and its extension to the full numerical model. All quantities in this subsection are at zero temperature and $\theta=45^\circ$, initially at a common fixed amplitude. In sums over retained phase sectors below, keeping only $\kappa=\pi$ gives the analytic model in End Matter; keeping $\kappa=0,\pi$ gives the full numerical theory. The limiting amplitude and response coefficients must be evaluated consistently for the chosen model. Define the signed, real gap form factors $d_{0,1}(\mathbf k)=\Delta_*w_{\mathbf k}\cos(2\alpha_{\mathbf k}\pm\pi/4)$ and $e_l=(\xi_{\mathbf k}^2+d_l^2)^{1/2}$. The determinant that enters the fermionic free energy is
\begin{equation}
\mathcal P(\omega,\mathbf k;\varphi)
=(\omega^2+e_0^2)(\omega^2+e_1^2)
+2t^2(\omega^2-\xi_{\mathbf k}^2+d_0d_1\cos\varphi)+t^4.
\end{equation}
The phase-dependent free energy is $-\intk\int d\omega\,\ln\mathcal P/(2\pi)$, up to phase-independent terms. Expanding its logarithm, the term proportional to $t^2\cos\varphi$ integrates to zero by the $45^\circ$ symmetry. The square of that term produces $\cos^2\varphi=(1+\cos2\varphi)/2$, giving $B=b_4t^4+o(t^4)$ with
\begin{align}
b_4
&=\intk\int_{-\infty}^{\infty}\frac{d\omega}{2\pi}
\frac{d_0^2d_1^2}{(\omega^2+e_0^2)^2(\omega^2+e_1^2)^2}=\intk\frac{d_0^2d_1^2(e_0^2+3e_0e_1+e_1^2)}
{4e_0^3e_1^3(e_0+e_1)^3}>0.
\label{S-eq:b4explicit}
\end{align}
The second line follows by differentiating $\int d\omega/[(\omega^2+e_0^2)(\omega^2+e_1^2)]/(2\pi)=1/[2e_0e_1(e_0+e_1)]$ with respect to $e_0^2$ and $e_1^2$. At an isolated node this coefficient is integrable because the numerator contains the square of the vanishing gap. The zero-temperature expansion is not uniform at the nodes, so the remainder is written as $o(t^4)$ without assuming an analytic higher-order series in $t^2$.

Let $r_s$ denote the physical trial Josephson mass $r_J$ on background $s=d,\chi$, not its spatial stiffness, and set $\sigma_d=+1$, $\sigma_\chi=-1$. To this order the variational energy is
\begin{equation}
F_s=F_{\mathrm{av}}+E_{G,s}(r_s;0,t)
+\sigma_sB e^{-2W_s}-\frac{r_sW_s}{2}+o(t^4),
\label{S-eq:weakFbranches}
\end{equation}
where $F_{\mathrm{av}}$ is the phase-independent Fourier component of the mean-field energy. It is the same for both backgrounds at a common amplitude, and $W_s$ is the physical phase variance on background $s$. The chiral stationary branch satisfies $r_\chi=4b_4Z_*t^4+o(t^4)$, while the nodal background is evaluated at the regularized zero-mass boundary, with negligible regulator contribution to the energy. Because $W_s(r_s;0,t)\to W_*$ as $t,r_s\to0$,
\begin{align}
E_{G,s}(r_s;0,t)-E_{G,s}(0;0,t)
&=\frac12\int_0^{r_s}dr\,W_s(r;0,t)=\frac{r_sW_*}{2}+o(r_s).
\end{align}
Thus the trial subtraction cancels the leading mass cost rather than the whole zero-point energy. The surviving splitting is
\begin{equation}
F_d-F_\chi=E_{G,d}(0;0,t)-E_{G,\chi}(0;0,t)+2b_4Z_*t^4+o(t^4).
\label{S-eq:CQsplit}
\end{equation}
This comparison takes the zero-temperature limit first and neglects the numerical floor in the energy; a fixed floor cannot resolve an arbitrarily small chiral mass.

The background term in Eq.~\eqref{S-eq:CQsplit} can be evaluated directly from the local response kernels. Define their phase differences at the common amplitude by
\begin{equation}
\rho_{\kappa4}=\lim_{t\to0^+}\frac{\rho_{\kappa,d}-\rho_{\kappa,\chi}}{t^4},
\qquad
c_{\kappa4}=\lim_{t\to0^+}\frac{\chi_{\kappa,d}-\chi_{\kappa,\chi}}{t^4}.
\label{S-eq:response4limits}
\end{equation}
The order-$t^2$ phase-sensitive terms are odd harmonics and cancel after angular integration. The first phase-selecting terms of the retained local kernels occur at order $t^4$. The finite limits in Eq.~\eqref{S-eq:response4limits} require node-resolved integration; they do not follow solely from the discrete phase symmetry. With $\rho_\kappa\to \rho_*$, $\chi_\kappa\to\chi_*$, and $K_\kappa(\mathbf q)=4N_z[\chi_*^{-1}+\widetilde V_\kappa(\mathbf q)]$, the massless zero-point energy and its difference give
\begin{align}
E_G(0;0,t)
&=\frac{1}{4\pi}\int_0^{\Lambda}d|\mathbf q|\,|\mathbf q|^2
\sum_{\kappa}\sqrt{\rho_\kappa K_\kappa(\mathbf q)},
\nonumber\\
g_4
&=\frac{1}{8\pi}\int_0^{\Lambda}d|\mathbf q|\,|\mathbf q|^2\sum_{\kappa}
\left[\sqrt{\frac{K_\kappa(\mathbf q)}{\rho_*}}\rho_{\kappa4}
-\frac{4N_z}{\chi_*^2}\sqrt{\frac{\rho_*}{K_\kappa(\mathbf q)}}c_{\kappa4}\right],
\nonumber\\
C_Q&=g_4+2b_4e^{-2W_*}.
\label{S-eq:CQexplicit}
\end{align}
The Coulomb interaction itself has no background phase dependence. Unlike $b_4$, the sign of $g_4$ is not fixed generally, and a chiral zero-temperature ground state requires $C_Q>0$.

For the relative-sector-only model in End Matter, the sign is positive for the microscopic parameters used here. Relaxing the amplitude at $t=0.25$ meV gives $\Delta_d\simeq35.4591$ meV. Holding this common amplitude and cutoff fixed, node-resolved integration gives $[E_{G,d}(0;0,t)-E_{G,\chi}(0;0,t)]/t^4\simeq(6.78,6.78,6.76)\times10^{-5}\,\mathrm{meV}^{-3}L_0^{-2}$ at $t=(0.25,0.5,1.0)$ meV, respectively, with $L_0=10$ nm. These values approach a positive weak-tunneling coefficient $g_4\simeq6.8\times10^{-5}\,\mathrm{meV}^{-3}L_0^{-2}$. The calculation uses cancellation-safe quasiparticle energies and composite Gauss--Legendre quadrature resolving the nodes. Increasing the rule from 40 to 48 points per subinterval changes the $t=0.25$ meV result by approximately $0.02\%$.

The following numerical checks retain both phase sectors and their own limiting amplitude. For the main-text parameters, they give $b_4\simeq4.67\times10^{-4}\,\mathrm{meV}^{-3}L_0^{-2}$, $e^{-2W_*}\simeq0.0504$, and a positive $g_4\simeq8.3\times10^{-5}\,\mathrm{meV}^{-3}L_0^{-2}$. Their sum gives $C_Q\simeq1.30\times10^{-4}\,\mathrm{meV}^{-3}L_0^{-2}$, consistent with the directly calculated $[F_d(0;t)-F_\chi(0;t)]/t^4$. For example, at $t=0.5$ meV and common amplitude $\Delta_d=29.6269$ meV, Eq.~\eqref{S-eq:CQsplit} gives $1.2963\times10^{-4}\,\mathrm{meV}^{-3}L_0^{-2}$, compared with $1.2964\times10^{-4}\,\mathrm{meV}^{-3}L_0^{-2}$ from the full free-energy difference.

Finally, consider amplitude relaxation. Write the zero-temperature energy as $F_s(\Delta_d;t)=F_{\mathrm{common}}(\Delta_d;t)+t^4h_s(\Delta_d)+o(t^4)$, where $h_s$ is the background-dependent coefficient. Let $\overline\Delta(t)$ minimize the common part, with limiting curvature $K_\Delta>0$. Then
\begin{align}
\Delta_s-\overline\Delta
&=-\frac{t^4\,\partial_{\Delta_d}h_s}{K_\Delta}+o(t^4),
\nonumber\\
F_s(\Delta_s;t)-F_s(\overline\Delta;t)
&=-\frac{t^8(\partial_{\Delta_d}h_s)^2}{2K_\Delta}+o(t^8).
\end{align}
Consequently the additional, state-dependent amplitude relaxation does not modify the leading $t^4$ splitting. The common amplitude tends to $\Delta_*$, at which the coefficients above are evaluated. This conclusion concerns the retained local-response SCHA and does not establish the same power law for a fully nonlocal or dissipative phase theory.

\subsection{Two-sector numerical benchmark for the low-temperature crossing}
\label{S-sec:two-sector-benchmark}
The overall-phase sector is not needed for the relative-phase mechanism in End Matter, but its free energy does not cancel identically between the nodal and chiral backgrounds. In the full calculation, nodal quasiparticles soften both physical stiffnesses. Their limiting depletion rate is $a_\rho=a_-=\mu\ln(2)/(4\pi\Delta_*)$ for $N_z=2$, using the same normalization as the main-text phase model. In the limit $t\to0^+$, the leading low-temperature slope of the full calculation is the sum of two positive contributions,
\begin{equation}
\Gamma^{\mathrm{full}}(0)=\frac{a_\rho}{8\pi\sqrt{\rho_*}}
\int_0^{\Lambda}d|\mathbf q|\,|\mathbf q|^2\left[\sqrt{K_0(\mathbf q)}+\sqrt{K_\pi(\mathbf q)}\right].
\label{S-eq:Gammafull}
\end{equation}
Here $\Delta_*$ and $\rho_*$ are the limiting amplitude and common spatial stiffness of the full theory. The $t\to0$ limit of Eq.~\eqref{eq:Gammasteps} in End Matter retains only the relative-sector contribution, written in terms of $\rho_*$ and $\chi_-$ using Eq.~\eqref{S-eq:physicalphasemap}, and uses the corresponding model's limiting amplitude. Adding the overall sector changes quantitative coefficients but is not required for the $t^4$ scaling.

Dedicated checks of the full local-response free energy, using node-resolved quadrature and cancellation-safe quasiparticle energies, give $\Gamma^{\mathrm{full}}(0)\simeq4.44/(10\,\mathrm{nm})^2$ and a crossing scale of approximately $T_\chi\simeq3.4\times10^{-4}\,\mathrm K\,[t/(1\,\mathrm{meV})]^4$. Direct crossings over $0.25\leq t\leq1.5$ meV give a fitted exponent of 3.99, approaching 4 for the weakest hoppings. The result is insensitive to practical changes of the mass floor and further amplitude relaxation. These numerical prefactors refer to the full two-sector calculation, not the relative-only analytic model.

\section{Phase diagrams and mean-field comparisons}
\label{S-sec:phase-comparisons}
\subsection{Tunneling--temperature comparison at \texorpdfstring{$45^\circ$}{45 degrees}}
\label{S-sec:tunneling-comparison}
\begin{figure}[htbp]
\centering
\includegraphics{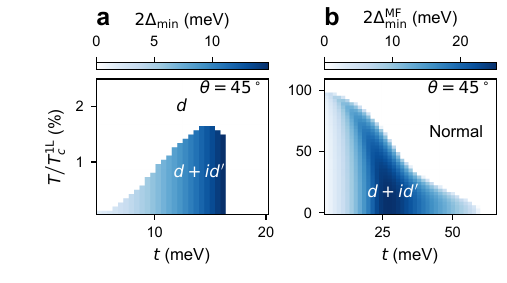}
\caption{\textbf{Mean-field pairing robustness does not ensure a high chiral ordering temperature.}
Minimum quasiparticle gap in the tunneling--temperature plane at $\theta=45^\circ$. Left and right panels include and omit phase fluctuations, respectively. Temperature is expressed as a percentage of $\TcL$, and white denotes zero minimum gap. The SCHA scan covers $5\leq t\leq20$ meV, while the mean-field reference extends to stronger tunneling. Independent temperature and color scales make explicit the separation between chiral ordering and pairing.}
\label{S-fig:tunneling-comparison}
\end{figure}

Figure~\ref{S-fig:tunneling-comparison} compares the two treatments at $\theta=45^\circ$. Mean-field chirality persists to the pairing transition throughout the range with a nonzero $d$-wave solution, and its excitation gap is largest near $t\simeq25$ meV. At very strong tunneling, pairing in the restricted $d$-wave model is suppressed. Disappearance of this solution around $t\gtrsim60$ meV is consistent with lattice-model analyses in which other pairing channels can become favorable~\cite{panda26gap}; those channels lie outside the present variational space.

The SCHA chiral scale is much lower and nonmonotonic. Within the sampled range $5\leq t\leq20$ meV, it remains below approximately $1.6\%$ of $\TcL$, and no chiral region is resolved at the displayed temperatures in the upper part of the scan, approximately 16 to 20 meV. The upper scan limit reflects the weak-tunneling local-response approximation and the range of the reported calculation, not an expansion of the microscopic Josephson energy in powers of $t$. The fluctuation-corrected panel is replotted in main-text Fig.~\ref{fig:phase-overview}(c).

The nodal spectrum gives a low-temperature interpretation of the turnover. On the nonchiral background $\varphi=0$, hold its zero-temperature amplitude $\Delta_d$ fixed and define $\zeta=2t^2/\Delta_d^2<1$. At $\theta=45^\circ$, Eq.~\eqref{S-eq:bands} gives nodes at $\xi_{\mathbf k}=0$ and $\cos(4\alpha_n)=-\zeta$, where $\alpha_n$ is a nodal polar angle. With $\xi=\xi_{\mathbf k}$ and $\delta\alpha=\alpha_{\mathbf k}-\alpha_n$, the lower branch has the expansion
\begin{equation*}
E_\pi^2\simeq\frac{1-\zeta}{1+\zeta}\xi^2
+4\Delta_d^2(1-\zeta)(\delta\alpha)^2.
\end{equation*}
Both nodal velocities decrease as $\zeta$ increases. Projecting the gradient vertices onto these cones leaves the overall-current vertex unchanged, while the squared relative-current vertex is reduced by $(1-\zeta)/(1+\zeta)$. Integrating their thermal occupation, using $n_F(E)=(e^{E/T}+1)^{-1}$ and $\int_0^\infty dE\,E[-\partial_E n_F(E)]=T\ln(2)$, gives, in the physical phase normalization,
\begin{align*}
\rho_{\kappa,d}(T)&=\rho_{\kappa,d}(0)-a_\kappa(t)T+o(T),
\qquad \kappa=0,\pi,\\
a_0(t)&=a_\Delta\frac{\sqrt{1+\zeta}}{1-\zeta},
\qquad a_\pi(t)=\frac{a_\Delta}{\sqrt{1+\zeta}},
\qquad a_\Delta=\frac{\mu\ln(2)}{4\pi\Delta_d}.
\end{align*}
Node-resolved evaluation of the retained kernels at $t\simeq14.68$ meV and $\Delta_d\simeq28.24$ meV reproduces these slopes within $0.002\%$. The density kernels have no linear-in-$T$ term. At fixed density response, trial mass, amplitude, and cutoff, the Gaussian zero-point contribution to the nodal free-energy slope is therefore
\begin{equation*}
\Gamma_G(t)=\sum_{\kappa=0,\pi}a_\kappa(t)
\left.\frac{\partial E_{G,d}}{\partial\rho_\kappa}
\right|_{\chi_0,\chi_\pi,r_J,\Lambda}>0.
\end{equation*}
The growing overall-sector depletion enhances the nodal state's thermal free-energy advantage, whereas chiral quasiparticles remain thermally inactive for $T\ll E_g$. Amplitude relaxation can reinforce this effect by reducing $\Delta_d$ and increasing $\zeta$. Thus the low-temperature guide $T_\chi\simeq\Delta f_0(t)/\Gamma(t)$ can turn downward even as the Josephson stabilization increases. Indeed, across the maximum in the stored scan at the lowest sampled temperature, $J_2$ increases and the chiral variance $W$ decreases; the turnover is not simply a weakening of $J_2$ or its Debye--Waller factor. This is a model-dependent interpretation, not a universal finite-$t$ law: the Dirac expansion is nonuniform near $\zeta=1$, where the nodes reconstruct, and the complete branch free energies, rather than an extrapolated divergent slope, determine the boundary.

\subsection{Angle--temperature phase diagrams}
\label{S-sec:angle-comparison}
\begin{figure}[htbp]
\centering
\includegraphics{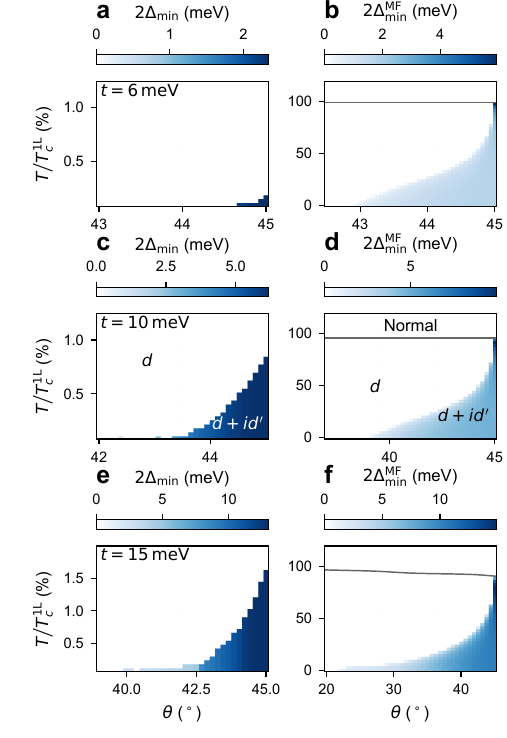}
\caption{\textbf{Fluctuations reduce the mean-field chiral window in angle and temperature.}
Minimum quasiparticle gap for $t=6$, 10, and 15 meV, from top to bottom. Left and right columns show SCHA and mean-field calculations, respectively. The diagrams are symmetric about $45^\circ$. Gray curves mark the mean-field normal--superconducting boundary where shown; both the normal and nodal superconducting states have zero minimum gap. Temperature is expressed as a percentage of $\TcL$. The two columns use different temperature ranges and color scales, with white denoting zero gap. The finite-gap region identifies the chiral background; the SCHA mass-floor qualification is the same as in main-text Fig.~\ref{fig:phase-overview}.}
\label{S-fig:phase-comparison}
\end{figure}

Figure~\ref{S-fig:phase-comparison} compares the fluctuation-corrected and mean-field phase diagrams for $t=6$, 10, and 15 meV. These maps extend the representative $t=10$ meV result in main-text Fig.~\ref{fig:phase-overview}(a). All temperatures in the comparison figures are expressed as percentages of the monolayer mean-field transition temperature $\TcL$, consistently with the main-text overview. The finite minimum gap identifies the chiral variational background in the weak-tunneling regime. A nodal background evaluated at the regularized zero-mass boundary should not be interpreted as proof of a coherent zero-phase state.

Mean-field theory gives a broad region of chirality extending to the superconducting pairing transition. Its low-temperature angular half-width grows from approximately $2^\circ$ at $t=6$ meV to approximately $23^\circ$ at $t=15$ meV; at $t=10$ meV it is about $6^\circ$. For $t=6$ and 10 meV, the mean-field transition temperature is nearly independent of twist and remains close to $\TcL$, consistent with the weak angle dependence of the junction transition reported experimentally~\cite{Zhao2023Science,Wang2023NatCommun}.

Phase fluctuations reduce both the angular and temperature ranges. At $t=6$ meV, the chiral region is nearly eliminated. At $t=10$ meV, it survives only within approximately $2^\circ$ of $45^\circ$ and below $T/\TcL\simeq0.8\%$. At $t=15$ meV, the low-temperature half-width is about $5^\circ$ and chirality remains below $T/\TcL\simeq1.6\%$. The large intralayer pairing scale therefore does not protect chirality over a comparable temperature interval.

The calculated gap jumps across the SCHA chiral boundary but evolves continuously across the mean-field boundary. This discontinuity is consistent with a first-order transition generated by the competition between distinct nodal and chiral free-energy minima. The calculation follows the largest positive trial-mass branch at every point; the jump is not produced by switching between distinct low- and high-mass roots of the same self-consistency equation.

\subsection{Nonmonotonic mean-field excitation gap near the pairing transition}
\label{S-sec:gap-temperature}
\begin{figure}[htbp]
\centering
\includegraphics{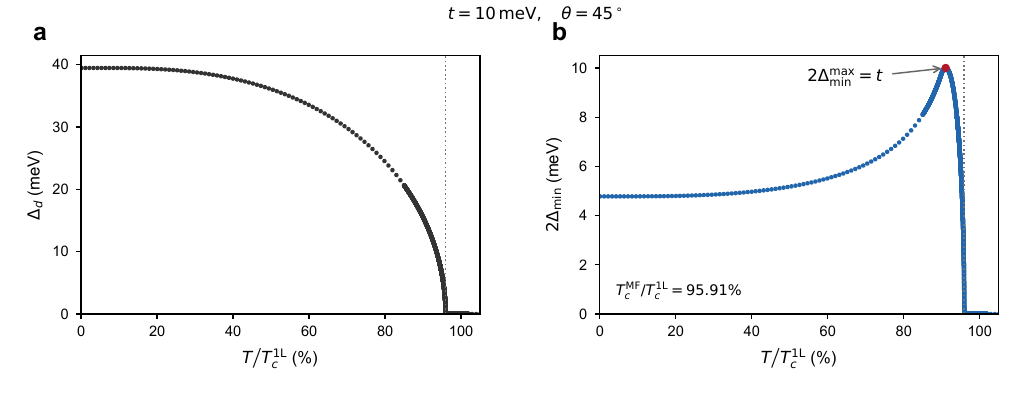}
\caption{\textbf{The mean-field excitation gap peaks below, and closes at, the pairing transition.}
Temperature cuts at $\theta=45^\circ$ and $t=10$ meV: (a) pairing amplitude $\Delta_d$ and (b) minimum quasiparticle gap $2\Delta_{\min}$. Dots show the full refined calculation without connecting lines, with denser sampling near the maximum and gap closure. The red point marks the maximum $2\Delta_{\min}=t$, reached at $T/\TcL=91.02\%$ and $\Delta_d=\sqrt{2}\,t$. Dotted vertical lines mark the bilayer pairing transition, $T_c^{\mathrm{MF}}/\TcL=95.91\%$, where both quantities vanish.}
\label{S-fig:gap-temperature}
\end{figure}

The enhanced color near the upper edge of the mean-field chiral region in Fig.~\ref{S-fig:phase-comparison}(d) reflects a genuine maximum below the pairing transition, not a finite excitation gap at that transition. Figure~\ref{S-fig:gap-temperature} resolves a temperature cut at $\theta=45^\circ$ and $t=10$ meV. The mean-field pairing amplitude $\Delta_d$ decreases monotonically, whereas the minimum positive quasiparticle energy $\Delta_{\min}=\min_{\mathbf k}E_\pi(\mathbf k)$ first increases and then vanishes continuously with $\Delta_d$. The plotted gap is $2\Delta_{\min}$.

This behavior follows directly from the microscopic spectrum. At $\theta=45^\circ$ and relative phase $\varphi=\pi/2$, Eq.~\eqref{S-eq:bands} inside the pairing shell becomes
\begin{equation}
E_\pi^2(\xi,\alpha)=\xi^2+t^2+\frac{\Delta_d^2}{2}
-\sqrt{\frac{\Delta_d^4}{4}\sin^2(4\alpha)+t^2(\Delta_d^2+4\xi^2)},
\label{S-eq:gap45spectrum}
\end{equation}
where $\xi=\xi_{\mathbf k}$ and $\alpha=\alpha_{\mathbf k}$. The angular minimum has $|\sin(4\alpha)|=1$. Minimizing over $\xi$ then gives $\xi_{\min}^2=\max\{0,t^2-\Delta_d^2/4-\Delta_d^4/(16t^2)\}$ and
\begin{equation}
2\Delta_{\min}=
\begin{cases}
\sqrt{\Delta_d^2+4t^2}-\Delta_d,
&\Delta_d\geq\Delta_\times,\\[2pt]
\Delta_d\sqrt{1-\Delta_d^2/(4t^2)},
&0\leq\Delta_d\leq\Delta_\times,
\end{cases}
\qquad
\Delta_\times=t\sqrt{2(\sqrt{5}-1)}.
\label{S-eq:gap45closed}
\end{equation}
For the parameters in Fig.~\ref{S-fig:gap-temperature}, the minimizing momenta remain inside the pairing shell; unpaired states outside it have higher energies. Thus $\Delta_{\min}\simeq t^2/\Delta_d$ for $\Delta_d\gg t$: decreasing the pairing amplitude initially increases the induced gap. Near the pairing transition, however, the minimizing momentum moves away from $\xi=0$ toward the normal-state bonding and antibonding Fermi surfaces, and $\Delta_{\min}\simeq\Delta_d/2\to0$. The exact maximum is $2\Delta_{\min}=t$ at $\Delta_d=\sqrt{2}\,t$.

The refined calculation gives this maximum at $T/\TcL=91.02\%$, followed by continuous closure at the bilayer mean-field pairing temperature $T_c^{\mathrm{MF}}/\TcL=95.91\%$. The latter is slightly below the monolayer reference $\TcL$. We resolve the final approach with a linear temperature spacing of approximately $0.0025\%$ of $\TcL$ and additional logarithmically spaced points down to $1-T/T_c^{\mathrm{MF}}=10^{-10}$. The amplitude is obtained from the derivative of the unexpanded mean-field free energy with respect to $\Delta_d^2$, with the normal-state derivative locating $T_c^{\mathrm{MF}}$. Doubling the radial and angular quadrature resolutions leaves the displayed curves unchanged; independent two-component minimizations with the production routine and momentum-space gap searches agree with the scalar amplitude solution and Eq.~\eqref{S-eq:gap45closed}, respectively. The original phase-diagram datasets are not modified.

\subsection{Large-tunneling mean-field comparison}
\label{S-sec:large-t-comparison}
\begin{figure}[htbp]
\centering
\includegraphics{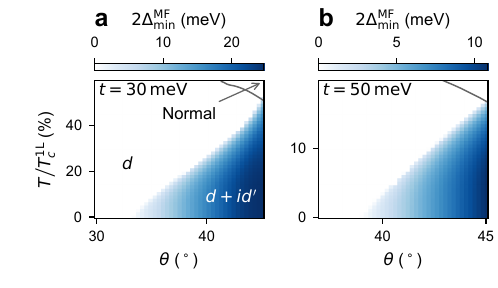}
\caption{\textbf{Large tunneling suppresses the pairing scale at large twist.}
Mean-field minimum-gap phase diagrams for $t=30$ and 50 meV. Temperature is expressed as a percentage of $\TcL$. The linear color scale starts at white for zero gap; gray curves distinguish the normal state from the nodal superconducting state, both of which have zero minimum gap. The strong angle dependence of the pairing transition contrasts with the weaker-tunneling reference in Fig.~\ref{S-fig:phase-comparison}.}
\label{S-fig:large-t}
\end{figure}

Figure~\ref{S-fig:large-t} shows the mean-field phase diagrams at $t=30$ and 50 meV. The transition temperature falls strongly as the twist approaches $45^\circ$, in contrast to the nearly angle-independent transition at smaller tunneling (Fig.~\ref{S-fig:phase-comparison}). This strong angle dependence differs from the junction measurements~\cite{Zhao2023Science,Wang2023NatCommun} and motivates using $t$ of order 6 to 10 meV as an effective-model reference~\cite{tummuru22josephson}. Increasing tunneling to make mean-field chirality more robust therefore also changes other observables. These panels are mean-field comparisons, not additional SCHA scans.

\section{Stiffnesses and relative-phase collective modes}
\label{S-sec:collective}
\subsection{Stiffnesses at the selected SCHA state}
\label{S-sec:selected-stiffness}
\begin{figure}[htbp]
\centering
\includegraphics{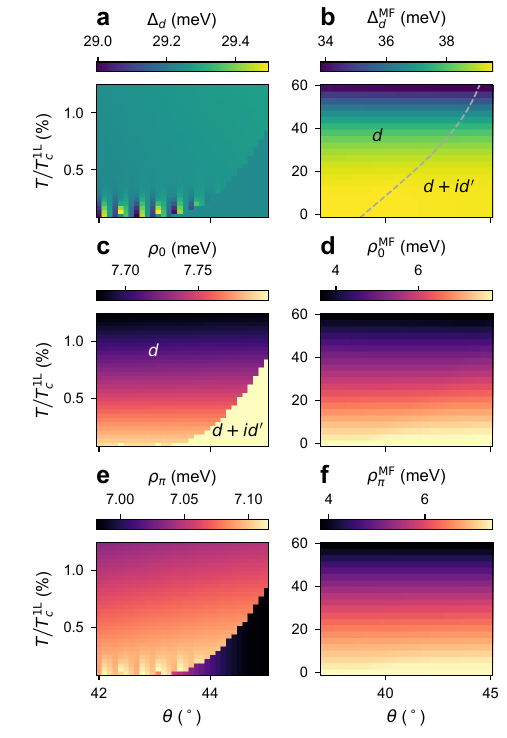}
\caption{\textbf{Pairing amplitudes and stiffnesses underlying the chiral reconstruction.}
Gap amplitude and overall and relative stiffnesses versus angle and temperature at $t=10$ meV. Left and right columns include and omit phase fluctuations, respectively. Temperature is expressed as a percentage of $\TcL$, as in main-text Fig.~\ref{fig:phase-overview}. Stiffnesses use the physical pair-phase normalization defined in Supplemental text~\ref{S-sec:methods}. The relative stiffness is the local gradient--gradient coefficient. All kernels are evaluated at the corresponding selected order parameter. The dashed curve identifies the mean-field chiral boundary.}
\label{S-fig:stiffness}
\end{figure}

Figure~\ref{S-fig:stiffness} compares gap amplitudes and spatial coefficients at $t=10$ meV. At low temperature, the mean-field amplitude is approximately 40 meV and the fluctuation-renormalized amplitude is approximately 30 meV. Mean-field $\rho_0$ and $\rho_\pi$ vary weakly across the displayed angular interval. With fluctuations, the overall stiffness is slightly larger in the chiral state than in the adjacent achiral state, whereas the relative stiffness shows the opposite contrast. Both vary little with temperature inside the narrow chiral region.

The response convention and its distinction from a physical helicity modulus are specified in Supplemental text~\ref{S-sec:microscopic}.

\subsection{Fluctuation and mean-field comparison of the Josephson plasma gap}
\label{S-sec:plasma-comparison}
\begin{figure}[htbp]
\centering
\includegraphics{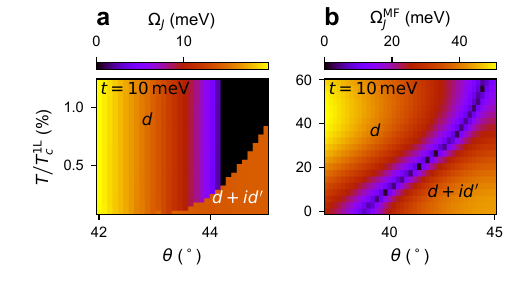}
\caption{\textbf{Phase fluctuations reconstruct the Josephson plasma gap.}
Relative-phase plasma energy in the twist-angle--temperature plane at $t=10$ meV, with phase fluctuations within SCHA (left) and at mean field (right). The SCHA panel uses the self-consistent mass $r_J$ and reproduces main-text Fig.~\ref{fig:phase-overview}(b); the mean-field panel uses the bare mass at its own equilibrium order parameter. Temperature is expressed as a percentage of $\TcL$. The panels have independent angular, temperature, and color scales; the mean-field temperature range is restricted to approximately $60\%$ of $\TcL$. At the zero-mass boundary, the small residual plotted energy is a numerical-regulator effect, not a physical plasma gap or a resolved locked solution.}
\label{S-fig:plasma-comparison}
\end{figure}

Figure~\ref{S-fig:plasma-comparison} compares the relative-phase plasma gap with and without phase fluctuations at $t=10$ meV. The SCHA panel reproduces main-text Fig.~\ref{fig:phase-overview}(b), while the mean-field panel covers a wider angular interval and temperatures up to approximately $60\%$ of $\TcL$, as in Fig.~\ref{S-fig:stiffness}. The SCHA energy follows $\Omega_J^2=r_J/\chi_-(\mathbf 0)$ with the self-consistent trial mass $r_J$. For the mean-field reference, the bare Josephson mass is $r_J^{\mathrm{bare}}=\partial_\varphi^2F_0$, evaluated at the mean-field minimum. Each calculation uses its own density susceptibility at the selected order parameter.

The mean-field relative-phase mode softens around the continuous time-reversal-symmetry-breaking boundary~\cite{tang26dynamical}. Within SCHA, strong softening extends over a region neighboring the surviving low-temperature chiral state, within approximately $0.8^\circ$ of $45^\circ$. The pairing amplitude varies much less over this region (Fig.~\ref{S-fig:stiffness}), distinguishing weak interlayer locking from loss of pairing. At $45^\circ$ above the chiral state, the mass floor serves as a numerical regulator of the zero-mass boundary, not a physical plasma gap or evidence for rigid $\varphi=0$ locking. Establishing the nature of the unlocked regime requires the additional correlation diagnostics discussed in the main article. The independent angular, temperature, and color scales in Fig.~\ref{S-fig:plasma-comparison} should be kept in mind when comparing the two treatments.

\subsection{Small-momentum dispersion of the relative mode}
\label{S-sec:smallq-dispersion}
The antisymmetric Coulomb potential has the expansion
\begin{equation}
\widetilde V_\pi(\mathbf q)
=\frac{e^2d}{2\epsilon_0\epsilon_b}
-\frac{e^2d^2}{4\epsilon_0\epsilon_b}|\mathbf q|+O(|\mathbf q|^2).
\end{equation}
Using Eq.~\eqref{S-eq:modesmain} gives
\begin{equation}
\left.\frac{1}{\hbar}\frac{d\omega_\pi}{d|\mathbf q|}\right|_{|\mathbf q|\to0^+}
=-\frac{N_ze^2r_J d^2}{2\hbar\epsilon_0\epsilon_b\Omega_J}.
\label{S-eq:vg}
\end{equation}
The negative initial group velocity arises from the momentum dependence of the interlayer electrostatic coupling. It is independent of $\rho_\pi$ at this order because the gradient contribution starts at $|\mathbf q|^2$. It vanishes without long-range Coulomb coupling or as the mass tends to zero, with magnitude proportional to $\sqrt{r_J}$ when the other coefficients remain finite. The overall phase mode remains gapless, $\omega_0\propto\sqrt{|\mathbf q|}$. For conventional multilayers, the leading Josephson harmonic often suffices to describe this structure~\cite{sun20collective,pokrovsky96spectroscopic,stinson14infrared,sun14indefinite,basov16polaritons,savelev10terahertz,gaifullin99caxis,sellati23generalized}; near $45^\circ$ in the present bilayer, the higher harmonics and their different fluctuation factors are essential.

This electrostatic mode dispersion is not a full spectroscopic response. Gabriele et al.~\cite{gabriele24charge} show that electromagnetic retardation mixes longitudinal and transverse plasma excitations in layered metals, modifying their small-momentum dispersion and density response. Sellati and Benfatto~\cite{sellati25ghost} show that counterflowing currents in a periodic bilayer with two Josephson couplings can hide a Josephson plasmon in the small-out-of-plane-momentum density response. Our isolated two-sheet model includes neither fully retarded electromagnetic dynamics nor probe-dependent spectral weights, so quantitative comparison with these different geometries requires an extended response calculation.

\end{document}